\documentclass[preprint,nofootinbib,superscriptaddress]{revtex4}

\usepackage{mathtools} 
\usepackage{bbold}
\usepackage{mathrsfs}
\usepackage{feynmp}	
\usepackage{subfigure} 

\begin{document}
\begin{fmffile}{graphs}

\title{When Does the Inflaton Decay?}

\author{C. Armendariz-Picon}
\affiliation{Department of Physics, St.~Lawrence University, Canton, NY 13617, USA}
\email{carmendarizpicon@stlawu.edu}

\begin{abstract}
In order for  the inflaton to decay into radiation at the end of inflation,  it needs to couple to light matter fields.   In this article we determine whether such couplings cause the inflaton to decay during inflation rather than after it. We calculate decay amplitudes during inflation, and determine to what extent such  processes have an impact on the  mean and variance of the inflaton, as well as on the expected energy density of its decay products.   Although the  exponential growth of the decay amplitudes with  the number of e-folds  appears to indicate the rapid decay of the inflaton, cancellations among different amplitudes and probabilities result in corrections to the different expectation values that only grow substantially when the number of e-folds is much larger than the inverse squared inflaton mass in units of the Hubble scale.  Otherwise, for typical parameter choices, it is safe to assume that the inflaton does not decay during inflation.

\end{abstract}
\maketitle

\section{Introduction}
In order for inflation to be successful, its end has to be followed by reheating, a period during which the universe is populated by radiation \cite{Kofman:1997yn,Allahverdi:2010xz,Amin:2014eta}.  Once this radiation thermalizes, the universe behaves as in the standard hot Big-Bang cosmology, and all its predictions,   from Big-Bang nucleosynthesis to the decoupling of the Cosmic Microwave Background are naturally reproduced and recovered. 

The simplest way to transfer the energy stored in the inflaton to that of radiation is for the inflaton to decay.  This is typically accomplished by coupling the inflaton itself to lighter matter fields, although gravitational particle production also appears to be possible \cite{Ford:1986sy,Damour:1995pd}. Yet once the inflaton is coupled to  matter, there is no guarantee that its decay will happen after the end of inflation, rather than during inflation. Naively, we would expect the inflaton to decay whenever its decay rate in flat spacetime $\Gamma$ is  larger than the Hubble constant $H$. Since the  inflaton decay rate is proportional to the square of a coupling constant,  at ``large" couplings the inflaton should decay during inflation, whereas  at  ``small" couplings  we would expect it to decay long after it. In fact, in a generally covariant theory the effective coupling constant that determines how rapidly the inflation decays  depends on  a positive power of the scale factor, which grows exponentially during inflation. As pointed out in \cite{Weinberg:2005vy,Weinberg:2006ac}, this suggests that quantum corrections may become large, and  thus spoil the standard inflationary predictions.

In this manuscript we attempt to determine whether the inflaton decays during inflation. The first question we need to face is what we  mean by ``decay." Even in flat spacetime, the definition of ``unstable particle" is not straightforward. But in the end, a particle is unstable when the probability for  a transition to a multi-particle final state (that of the decay products) is non-zero. Such transition  probabilities are in fact what determines  the decay rate of the inflaton during reheating \cite{Kofman:1997yn,Allahverdi:2010xz,Amin:2014eta}. Yet, as we shall see,   matters are not as simple during inflation. Here  we mostly consider  three measures that we believe capture the concept of decay: i) The  probabilities for the inflaton state to evolve into various multi-particle states, ii) the expectation value of the energy-momentum tensor of the inflaton and its decay products, and iii) the corrections that the couplings to matter introduce in the evolution of the inflaton. The first  is the analogue of what is calculated in flat spacetime and in perturbative studies of reheating.    The second quantifies the backreaction of the inflaton decay products on the universe's expansion, and the third directly addresses the impact on the evolution of the background (classical) inflaton field. 

One of the  main obstacles we encounter in the calculation of these measures is the appearance of divergences in the integrals over the modes the inflaton couples to. In flat spacetime, rigorous theorems  guarantee that  these divergences can be appropriately removed by renormalization of coupling constants and fields, and there are well-defined algorithms that detail how to do so \cite{BPHZ}.  But in  curved spacetimes renormalization is   much less developed.  In this manuscript we adopt  adiabatic subtraction as a regularization and renormalization scheme \cite{Parker:1974qw,Birrell:1982ix}. The main advantages of this scheme is that it can be directly implemented in a time-dependent background, without the need to formulate the theory in a manifestly covariant way, that it accomplishes regularization and renormalization in one fell swoop, and that it is one of the main schemes used in the literature on the topic, particularly in the context of calculations of the expectation value of the energy-momentum tensor. As a result, the implementation of adiabatic subtraction in a cosmological setting is relatively simple and straightforward.  The main disadvantages of the scheme are that manifest covariance is lost, and that the connection with the counterterms in the action is neither obvious nor manifest. Although adiabatic subtraction has been successfully applied in free field theories,  it also remains unclear to what extent it is justified once interactions are included \cite{Markkanen:2013nwa}. But, overall, these shortcomings just reflect the status of the renormalization program in curved spacetimes, rather than those of the adiabatic scheme itself. 

There is a huge literature on the effect of quantum corrections on the evolution of the inflaton (see for instance \cite{Herranen:2016xsy} and references therein), but, to our knowledge, there are not many references that investigate the matter in the context of the inflaton's decay during inflation.  Our work somewhat overlaps with reference \cite{Boyanovsky:2004ph}, whose  methods and focus on the inflaton fluctuations significantly deviate from  our analysis. It is also related to articles  that study axion-like couplings of the inflaton to gauge fields, such as \cite{Anber:2009ua}, although the nature of these couplings, and the evolution of the matter fields are very different from what we consider here.    Whether the inflaton decays or not during inflation may also have  implications for  the warm inflation scenario initially proposed in \cite{Berera:1995ie}. In the latter, the inflaton is assumed to decay during inflation, and the resulting  radiation is argued to modify the dynamics of the inflaton and somehow prolong the duration of inflation. Although warm inflation is not our main focus, our results could be used to check its underlying assumptions.

\section{Action}

Our main goal is to study how  the  couplings of the inflaton needed to reheat the universe affect the background dynamics during inflation.  We model the inflaton as an homogeneous scalar field $\phi$, and its  decay products as a single, massless scalar $\chi$,
\begin{equation}\label{eq:S}
	S=\int d^4x \sqrt{-g}
	\left[
		-\frac{1}{2}\partial_\mu\phi \partial^\mu \phi
		-\frac{1}{2}m_\phi^2 \phi^2
		-\frac{1}{2}\partial_\mu\chi \partial^\mu \chi
		-\frac{\lambda}{2} \phi \chi^2\right].
\end{equation} 
We assume that the inflaton potential is quadratic because it is then simpler to identify quantum states  that behave classically, although, actually, most of our results  hold for an arbitrary potential.   We expect the dominant couplings of the inflaton to be captured by  renormalizable terms, namely, cubic and possibly quartic couplings if we model matter by a scalar. These are in in fact the couplings that have been mostly considered in the literature \cite{Kofman:1997yn}.   Note that we do not include  any counterterms in the action,  nor  any of  the additional possible operators compatible with the symmetries of the theory (general covariance and a $\chi\to-\chi$ symmetry). In the adiabatic subtraction scheme the divergences are removed directly from the corresponding mode integral, without the need to invoke any counterterms.

Leaving gravity aside, the only interaction in the action (\ref{eq:S}) consists of the cubic coupling proportional to $\lambda$. Our aim is to work in perturbation theory, so we shall assume that $\lambda$ is ``small." Since $\lambda$ has dimensions of mass, it may not be clear what this means at first. In flat spacetime, at one loop, the renormalized effective potential for $\phi$  that follows from the action (\ref{eq:S}) is \cite{Coleman:1973jx}
\begin{equation}\label{eq:Veff}
	V_\mathrm{eff}(\phi)=\frac{1}{2}m_\phi^2 \phi^2
	+\frac{\lambda^2 \phi^2}{64\pi^2} \log \frac{\lambda\phi}{\phi^2_\mu},
\end{equation}
where $\phi_\mu$ is a renormalization scale.  Because  quantum corrections are small whenever the logarithmic term is subdominant, for the time being it appears that $\lambda$ is small whenever $\lambda\ll m_\phi$.  Since gravitational couplings are suppressed by $1/M_P$, we expect gravitational corrections to be much smaller than those induced by  renormalizable couplings, such as the one proportional to $\lambda$.  

\subsection{Hamiltonian}
In order to obtain a manifestly unitary time evolution, we need to find the Hamiltonian of the theory first. 
Since we are interested in inflation, we shall consider  a  spatially flat FRW universe,
\begin{equation}
	ds^2=a^2(t)\left[-dt^2+d\vec{x}^2\right],
\end{equation}
where the time coordinate $t$ is conformal time.  In that case the canonical momenta are
\begin{equation}\label{eq:pi}
	\pi_\phi=a^2 \dot{\phi}, 
	\quad
	\pi_\chi=a^2 \dot{\chi},
\end{equation}
and, therefore,  the Hamiltonian reads
\begin{equation}\label{eq:HT}
	\mathscr{H}=\int d^3x \left[
	\frac{\pi_\phi^2}{2a^2}
	+\frac{a^2}{2} (\vec{\nabla}\phi)^2
	+\frac{a^4 m_\phi^2}{2}\phi^2
	+\frac{\pi_\chi^2}{2a^2}
	+\frac{a^2}{2} (\vec{\nabla}\chi)^2
	+\frac{a^4 \lambda}{2}\phi\chi^2
	\right].
\end{equation}

We shall later be interested in quantities like the expectation value of the inflaton field in Fourier space, $\langle \phi(t,\vec{k})\rangle$. In an infinite universe the latter is proportional to $\delta(\vec{k})$  and therefore diverges for the zero mode $\vec{k}=0$. It is hence convenient to perform a  transformation to a set of discrete canonical fields in a finite volume universe, in which such expectation values remain finite.  We  assume that our fields live in a finite universe of comoving volume $V=L^3$ and impose periodic boundary conditions on the latter. Then the fields can be expanded as
\begin{subequations}
\begin{align}
	\phi=\frac{1}{\sqrt{V}}\sum_{\vec{k}}\phi_{\vec{k}}(t) e^{i\vec{k}\cdot\vec{x}},
	\quad 
	\pi_\phi=\frac{1}{\sqrt{V}}\sum_{\vec{k}} \pi^\phi_{\vec{k}}(t)e^{-i\vec{k}\cdot\vec{x}},
	\\
	\chi=\frac{1}{\sqrt{V}}\sum_{\vec{k}} \chi_{\vec{k}}(t)e^{i\vec{k}\cdot\vec{x}},
	\quad 
	\pi_\chi=\frac{1}{\sqrt{V}}\sum_{\vec{k}} \pi^\chi_{\vec{k}}(t)e^{-i\vec{k}\cdot\vec{x}},
\end{align}
\end{subequations}
where the sums run over 
$
	\vec{k}=\frac{2\pi}{L} \vec{n}, 
\,
	\vec{n}\in \mathbb{Z}^3,
$
and the Fourier modes satisfy the  Poisson bracket relations
\begin{equation}
	\{\phi_{\vec{k}},\pi^\phi_{\vec{k}'}\}=\delta_{\vec{k}\vec{k}'},
	\quad
	\{\chi_{\vec{k}},\pi^\chi_{\vec{k}'}\}=\delta_{\vec{k}\vec{k}'}.
\end{equation}
Note that $\pi^\phi_{\vec{k}}$ is the canonical momentum conjugate to $\phi_{\vec{k}}$. The introduction of a finite volume universe is also useful to regularize the infrared divergences that we shall encounter below. 

Translational invariance implies that only the $\vec{k}=0$ mode $\phi_0$ can have a non-vanishing expectation value. We can isolate  this mode by averaging the inflaton over the whole universe,  
\begin{equation}
\phi_V\equiv \frac{1}{V} \int_V d^3 x \, \phi(\vec{x})=\frac{\phi_0}{\sqrt{V}}.
\end{equation}
Because of the explicit volume factor, the expectation value of the inflaton is not that of its zero mode $\phi_0\equiv \phi_{\vec{k}=0}$, but instead
\begin{equation}\label{eq:zero mode}
	\bar{\phi}\equiv\langle\phi\rangle=\langle \phi_V\rangle=\frac{\langle \phi_0\rangle}{\sqrt{V}}\equiv\frac{\bar\phi_0}{\sqrt{V}}.
\end{equation}
Our goal is to study the evolution of the homogeneous inflaton, so it suffices to focus on the evolution of the zero mode $\phi_0$. The restriction of the Hamiltonian (\ref{eq:HT}) to this mode results in
\begin{equation}\label{eq:H zero mode}
	\mathscr{H}= 
	\frac{(\pi^\phi_0)^2}{2a^2}
	+\frac{m_\phi^2 a^4}{2}\phi_0^2
	+
	\sum_{\vec{k}}
	\left[
	\frac{\pi^\chi_{\vec{k}}\, \pi^\chi_{-\vec{k}}}{2a^2}
	+\frac{a^2 k^2}{2} \chi_{\vec{k}}\chi_{-\vec{k}}
	+\frac{a^4\lambda}{2\sqrt{V}}\phi_0\,\chi_{\vec{k}}\chi_{-\vec{k}}
	\right].
\end{equation}
The zero mode $\phi_0$ thus couples to  $\chi$ through the cubic, momentum-conserving interaction $\phi_0\,\chi_{\vec{k}}\chi_{-\vec{k}}$. Setting the coupling $\lambda$ to zero in equation (\ref{eq:H zero mode}) we recover the free Hamiltonian of the theory, 
\begin{equation}\label{eq:H0}
	\mathscr{H}_0= 
	\frac{(\pi^\phi_0)^2}{2a^2}
	+\frac{m_\phi^2 a^4}{2}\phi_0^2
	+
	\sum_{\vec{k}}
	\left[
	\frac{\pi^\chi_{\vec{k}}\, \pi^\chi_{-\vec{k}}}{2a^2}
	+\frac{a^2 k^2}{2} \chi_{\vec{k}}\chi_{-\vec{k}}
	\right].
\end{equation}

\section{Quantization}
The classical equations of motion of the model admit a solution along which the inflaton slowly rolls down its potential, while the matter fields remain equal to zero. Yet in the quantum theory, we cannot set $\chi=0$ because of the zero point  fluctuations. Since the inflaton couples to $\chi^2$, such vacuum fluctuations end up modifying  the  evolution of the zero mode inflaton. Alternatively, we can think of the such couplings as inducing the decay of the inflaton into matter field quanta.

In order to determine the impact of quantum corrections on the background evolution, we obviously need to quantize the theory. We shall treat the inflaton interactions perturbatively by resorting to the interaction picture. In the latter, the fields follow the time evolution determined by the free Hamiltonian, which we study first. 

\subsection{Free Fields}
\label{sec:Free Fields}
In the free quantum theory, the Heisenberg  operators $\phi_0(t)$ and $\pi^\phi_0(t)$ satisfy the equal-time commutation relations $[\phi_0(t),\pi^\phi_0(t)]=i$ and the Heisenberg equations
\begin{subequations}
\begin{align}
	i\dot{\phi_0}=[\phi_0,\mathscr{H}_0]=i \frac{\pi^\phi_0}{a^2},
	\\
	i\dot{\pi}^\phi_0=[\pi^\phi_0,\mathscr{H}_0]=-i a^4 m_\phi^2 \phi_0,
\end{align}	
\end{subequations}
where $\mathscr{H}_0$ is the free (quadratic) Hamiltonian (\ref{eq:H0}).    In order to find a solution of these equations, we expand $\phi_0$ and $\pi_0$    in creation and annihilation operators $b^\dag$ and $b$ as usual,
\begin{equation}\label{eq:creation annihilation}
	\phi_0= u(t)b+u^*(t) b^\dag,
	\quad
	\pi^\phi_0= a^2[\dot{u}(t)b+\dot{u}^*(t) b^\dag],
\end{equation}
where $[b,b^\dag]=1$. The  time dependent coefficients $u(t)$ are determined by the condition that the fields satisfy the Heisenberg equations and the canonical commutation relations. Thus, $u$ needs to obey the equation of motion
\begin{equation}\label{eq:u motion}
	\ddot{u}+2\mathcal{H}\dot{u}+m_\phi^2 a^2 u=0,
\end{equation}
where $\mathcal{H}=\dot{a}/a$,
and satisfy  the normalization condition
\begin{equation}
	a^2[u\dot{u}^*- u^*\dot{u}]=i.
\end{equation}
That $a^2[u\dot{u}^*-u^* \dot{u}]$ is constant follows from equation (\ref{eq:u motion}), which also happens to be the field equation satisfied by the background inflaton solution.

The quantization of the matter fields $\chi_{\vec{k}}$  proceeds along the same lines, with the minor difference that, in order for $\chi_{\vec{k}}$ to carry a well-defined momentum, the  creation and annihilation operators must involve opposite momenta,
\begin{equation}
	\chi_{\vec{k}}=w_k(t) c_{\vec{k}} +w_k^*(t) \, c^\dag_{-\vec{k}},
	\quad
	\pi^\chi_{\vec{k}}=a^2[\dot{w}_k c_{-\vec{k}}+\dot{w}_k^* c^\dag_{\vec{k}}].
\end{equation}
In these expressions we have used isotropy, namely, that the equation of motion satisfied by $w_{\vec{k}}\equiv w_k$ only depends on the magnitude of $\vec{k}$, as we shall see below.  As before, in order for $\chi_{\vec{k}}$ and $\pi^\chi_{\vec{k}}$ to satisfy the canonical commutation relations, the $c_{\vec{k}}$ and $c^\dag_{\vec{k}}$ must satisfy the commutation relations $[c_{\vec{k}},c^\dag_{\vec{k}'}]=\delta_{\vec{k}\vec{k}'}$ and the modes $w_k$ must be properly normalized,
$
	a^2[w_k\dot{w}_k^*-w_k^* \dot{w}_k]=i.
$

\subsection{Coherent States}

We would like the  quantum state of the inflaton to reproduce  the properties of a classical rolling scalar field. In the case of the harmonic oscillator, states with classical properties are known as ``coherent states," and are defined to be eigenvectors of the annihilation operator. Since the free Hamiltonian of the inflaton (\ref{eq:H0}) resembles that of an harmonic oscillator,  we shall choose the state of the inflaton in analogy with such coherent  states, 
$
{|\beta\rangle\equiv N \exp(\beta\, b^\dag)|0\rangle,
}
$
 where $\beta$ is a constant that characterizes the state, $N$ is a normalization factor and $b|0\rangle=0$. 
 
 The expectation value of  the inflaton field $\phi$ in  such a coherent state is
\begin{equation}\label{eq:bar phi}
	\bar{\phi}=\frac{1}{\sqrt{V}}\left[\beta u(t)+\beta^* u^*(t)\right],
\end{equation}
which is the same as that of a  rolling  inflaton with  field value $2\mathrm{Re}[\beta u]/\sqrt{V}$. In particular, $\bar{\phi}$ satisfies the  equation of motion of a classical scalar field in an expanding universe. The variance of the averaged inflaton  in a coherent state is
\begin{equation}
\langle\Delta\phi_V^2\rangle\equiv \langle (\phi_V-\bar\phi)^2\rangle=\frac{u^*u}{V},
\end{equation}
which can be interpreted as  stating that the variance of $\phi_V$ is of order of the power spectrum on scales of the size of the universe.  As long as $\langle\Delta\phi^2_V\rangle\ll \bar\phi^2$  the expectation of the inflaton thus behaves like a classical scalar field. In view of equation (\ref{eq:bar phi}) this is satisfied for sufficiently large $\beta\gg 1$, in analogy with the classical limit of the harmonic oscillator.  Here we are  dealing in fact with a two-parameter class of coherent states: The parameter $\beta$ determines the field expectation, and the overall magnitude of $u$ determines its variance. Note that, as defined, $\langle\Delta\phi_V^2\rangle$ is quite different from $\langle \Delta\phi^2\rangle$. The former is just the variance of the zero mode alone, whereas the latter is the sum of  variances of all the modes, $\langle \Delta\phi^2\rangle=(1/V)\sum_{\vec{k}} \langle \phi_{\vec{k}} \phi_{-\vec{k}}\rangle.$

We choose the state of the matter fields to be annihilated by the operators $c_{\vec{k}}$, $c_{\vec{k}}|0\rangle=0$.  In the free theory, at tree level,  this implies that the expectation of $\chi$ is zero, $\langle \chi\rangle=0$. In other words, in the classical theory the matter fields $\chi$ vanish. In the quantum theory $\langle \chi\rangle$ remains zero because of the $\chi\to-\chi$ symmetry of the theory.

\subsection{Shifted Inflaton}

Because the expectation of $\phi$ in a coherent state is non-zero, these states are not particularly convenient for perturbative calculations, which largely rely on Feynman rules that assume vanishing $\langle \phi\rangle$. It is thus convenient to shift the canonical pair $(\phi_0$, $\pi^\phi_0$) by its expectation $\bar{\phi}_0$ and $\bar\pi^\phi_0$, 
\begin{equation}\label{eq:shifted field}
	\phi_0\equiv \bar{\phi}_0+\Delta\phi_0,
	\quad
	\pi^\phi_0\equiv \bar{\pi}^\phi_0+\Delta\pi^\phi_0.
\end{equation}
Since coherent states have Gaussian wave functions, it is easy to see, at least perturbatively,\footnote{A simple, though formal, proof can be obtained by changing variables in  the generating functional  for the $n$-point functions of $\phi_0$, $Z[J(t)]\equiv \int D\phi_0(t)
\exp(iS_2[\phi_0]) \exp(i \int J(t) \phi_0(t))$.} that working with a field $\phi_0$ in a coherent state with $\langle \beta| \phi_0|\beta\rangle=\bar{\phi}_0$ is mathematically equivalent to working in a theory with a shifted field $\Delta\phi_0$ in a state $|0\rangle$ with $\langle 0|\Delta\phi_0|0 \rangle=0$ and correlation
\begin{equation}
	\langle 0  |
		\Delta\phi_0(t_1)\Delta\phi_0(t_2)
			|0\rangle
			=u(t_1) u^*(t_2). 
\end{equation}
The latter are in fact the relations we would obtain by expanding $\Delta \phi_0$ and $\Delta\pi_0^\phi$ as in  equation (\ref{eq:creation annihilation}), with the same mode functions $u$. 

Shifting the inflaton by its expectation somewhat changes the structure of the Hamiltonian. In terms of the shifted field, the  Hamiltonian of the theory (\ref{eq:H zero mode}) becomes
\begin{equation}\label{eq:H delta zero mode}
	\mathscr{H}= 
	\frac{(\Delta\pi^\phi_0)^2}{2a^2}
	+\frac{m_\phi^2 a^4}{2}\Delta\phi_0^2
	+
	\sum_{\vec{k}}
	\left[
	\frac{\pi^\chi_{\vec{k}}\, \pi^\chi_{\vec{-k}}}{2a^2}
	+\frac{a^2 k^2}{2} \chi_{\vec{k}}\chi_{-\vec{k}}
	+\frac{a^4\lambda}{2\sqrt{V}}(\bar{\phi}_0+\Delta\phi_0) \,\chi_{\vec{k}}\chi_{-\vec{k}}\right],
\end{equation}
where we have used that $\bar{\phi}_0$ and $\bar\pi^\phi_0$ satisfy the free Hamiltonian equations (note that equations (\ref{eq:shifted field}) define a canonical transformation.) In perturbation theory in $\lambda$ we have the freedom to regard the terms proportional to $\lambda \bar{\phi}_0\chi_{\vec{k}}\chi_{-\vec{k}}$ as part of the interaction, or as part of the free theory. The latter is a better approximation, so we shall choose the free Hamiltonian to be  
\begin{equation}
\mathscr{H}_0= 
	\frac{(\Delta\pi^\phi_0)^2}{2a^2}
	+\frac{m_\phi^2 a^4}{2}\Delta\phi_0^2
	+
	\sum_{\vec{k}}
	\left[
	\frac{\pi^\chi_{\vec{k}}\, \pi^\chi_{\vec{-k}}}{2a^2}
	+\frac{a^2 k^2}{2} \chi_{\vec{k}}\chi_{-\vec{k}}
	+\frac{a^4\lambda}{2\sqrt{V}}\bar{\phi}_0\,\chi_{\vec{k}}\chi_{-\vec{k}}
	\right].
\end{equation}
At this point it is important to notice that, once the transition to the shifted field $\Delta\phi_0$ has been accomplished, our results become applicable to any inflationary potential $V(\phi)$, provided that we simply identify 
\begin{equation}
	m_\phi^2\equiv\frac{d^2 V}{d\phi^2}\equiv V_{\phi\phi}.
\end{equation}
What is special about the quadratic potential is the absence of any inflaton self-couplings. As long as the latter are weaker than the couplings to matter, our analysis should apply without modifications. 

\subsection{Mode Functions}

By assumption, the  background field $\bar{\phi}$   satisfies  the  equation of motion of a scalar in an expanding universe. For simplicity we are going to look at solutions of  the background equations in the limit in which the slow roll parameter $\epsilon\equiv \dot{\mathcal{H}}/\mathcal{H}^2-1$ approaches zero and the universe expands as in de Sitter space, and also  in the limit in which $\eta\equiv V_{\phi\phi}/H^2$  tends to zero and the inflaton field remains frozen (modulo a decaying solution),
\begin{subequations}\label{eq:background}
\begin{align}
	a&=-\frac{1}{Ht},\label{eq:a}\\
 \bar{\phi}_0&=const. 
\end{align}
\end{subequations}
Recall that  in de Sitter, conformal time $t$ extends from $t=-\infty$ to $t=0$, and thus remains negative throughout history.  Although it is essential to consider deviations from de Sitter when discussing inflationary perturbations, in our context there is not much loss of generality in the de Sitter limit. On the other hand, the simple structure of equations (\ref{eq:background}) will simplify many of our analytical calculations considerably.

The  normalized solution of the mode equation (\ref{eq:u motion}) in the limit $\epsilon\to0$ can be taken to be
\begin{subequations}
\begin{equation}
	u(t)=\frac{1}{a}\frac{1}{\sqrt{2k_L}}
	\left(-k_L t\right)^{1/2-r}
	\left[1+\frac{i}{2r}\, \left(-k_L t\right)^{2r}\right],
\end{equation}
where $k_L$ is an arbitrary constant with dimensions of inverse length, and we have defined
\begin{equation}
	r=\sqrt{\frac{9}{4}-\eta} , \quad \eta\equiv \frac{V_{\phi\phi}}{H^2}.
\end{equation}
\end{subequations}
In the limit $\eta\to 0$, the mode function $u(t)$ approaches a constant modulo a decaying term, just  like the background solution  $\bar\phi$,
\begin{equation}\label{eq:u limit}
	u(t)\to \frac{H}{\sqrt{2k_L^3}} \left[1+\frac{i}{3}\left(-k_Lt\right)^{3-\eta/3}\right].
\end{equation}
Although it can often be neglected, in some cases cancellations force us to keep track of the decaying mode. In those instances, time integrals often diverge as $\eta\to 0$, which is why we keep a non-zero $\eta$ in the exponent of the decaying mode. 

The value of $k_L$ has remained arbitrary so far. By its very nature, the zero mode always remains outside the horizon, so there is no way to determine its amplitude by tracing its evolution back to the short-wavelength limit during inflation. We shall instead assume that our finite volume universe is part of a larger inflationary patch, so that the mean square fluctuation of the scalar on scales of our finite universe is what one expects  from inflation, namely, about  $H^2$.   Since the mean square fluctuation of the scalar on scales of the volume of the universe is given by
\begin{equation}\label{eq:power}
	\langle\Delta\phi_V^2\rangle= \frac{|u|^2}{V} \approx \frac{1}{V}\frac{ H^2}{2k_L^3}\left(1+\frac{(k_L t)^6}{9}\right).
\end{equation}
we shall simply set $k_L=1/L$ and assume that the comoving size of the finite universe $L$ is much larger than the comoving horizon, $-k_L t\ll1$.   In fact, $u$ in equation (\ref{eq:u limit}) has the structure of the mode function of a light massive field in de Sitter space in the long-wavelength limit, provided we identify the wave number of the mode with $k_L$.  In that sense, we can think of $k_L$ as the wave number of our finite universe, and of $(-k_L t)^{-1}\equiv e^{N_L}$ as the exponential of the number of e-folds since that mode left the horizon. For the same reason, we should not trust the form of $u$ in the regime $-k_L t\gtrsim 1$. In order to keep our zero mode normalization explicit, we shall  keep all the factors of $k_L$ in our equations. For a given arbitrary variance of the zero mode, the value of $k_L$ is then determined by  equation (\ref{eq:power}).

With the inflaton shifted by its tree-level expectation, the equation of motion for the matter field mode functions becomes
\begin{equation}\label{eq:w motion}
	\ddot{w}_k+2\mathcal{H}\dot{w}_k+\left(k^2+\frac{\lambda \bar\phi_0}{\sqrt{V}}\, a^2\right) w_k=0.
\end{equation}
The coupling to  the inflaton has introduced an effective mass for the field $\chi$,   
\begin{equation}\label{eq:meff}
	m_\chi^2\equiv \frac{\lambda\bar \phi_0}{\sqrt{V}}=\lambda \bar{\phi}.  
\end{equation}
In the de Sitter limit we can  readily solve  for the mode functions of matter $w_k$, since the effective mass $m_\chi=\lambda \bar \phi$ remains essentially constant. In that case the mode functions are 
\begin{subequations}
\begin{equation}\label{eq:wk nu}
	w_k(t)=\frac{1}{a}\frac{\sqrt{-\pi t}}{2} \exp\left(\frac{i\pi \nu}{2}\right)H_1\left(\nu,-kt\right)\equiv \frac{\sqrt{-t/2}}{a} v(-kt),
	\quad
	\nu\equiv \sqrt{\frac{9}{4}-\frac{\lambda \bar\phi}{H^2}},
\end{equation}
where $H_1$ is the Hankel function of the first kind.
We have included an apparently irrelevant phase in the mode function because for a sufficiently  massive field, $\lambda\bar\phi> 9H^2/4$, the order of the Hankel function becomes imaginary, $\nu=i\mu$. In this case the mode functions are
\begin{equation}
w_k(t)=\frac{1}{a}\frac{\sqrt{-\pi t}}{2}\exp\left(-\frac{\pi \mu}{2}\right)H_1\left(i \mu,-kt\right)\equiv \frac{\sqrt{-t/2}}{a} v(-kt),
\quad
\mu=\sqrt{\frac{\lambda\bar\phi}{H^2}-\frac{9}{4}}.
\end{equation}
\end{subequations}
 In the limit of light matter fields, $m_\chi^2\ll H^2$ the mode functions (\ref{eq:wk nu}) approach those of a massless field,
\begin{equation}\label{eq:w massless infl}
	w_k=\frac{1}{a} \frac{e^{-i k t}}{\sqrt{2k}}\left(1-\frac{i}{k t}\right). 
\end{equation}

\subsection{Interactions}

In order to determine the evolution of the inflation in the presence of interactions, we need to solve the Heisenberg equations in the full interacting theory. Because this is not feasible, we resort instead to perturbation theory in the interaction picture. In this approach, operators $\mathcal{O}$ carry the free time evolution,
\begin{equation}
	i\frac{d\mathcal{O}_I}{dt}=[\mathcal{O}_I(t),\mathscr{H}_0(t)]+\frac{\partial{\mathcal{O}_I}}{\partial t},
\end{equation}
and states evolve with the interaction Hamiltonian. At time $t$ the state of the system is
\begin{equation}
	|\psi(t)\rangle=U_I(t, -T)|\psi(-T)\rangle,
\end{equation}
where $-T$ is the time at which the interaction picture is introduced, and 
 \begin{equation}\label{eq:UI}
 	U_I(t, -T)=\mathcal{T}\exp\left(-i\int_{-T}^t \mathscr{H}_I(t_1) \, dt_1\right)
 \end{equation}
is the time evolution operator in the interaction picture. As usual $\mathcal{T}$ is the time-ordering operator, and $\mathscr{H}_I$ is the interaction Hamiltonian in the interaction picture. In the case at hand, from equation (\ref{eq:H delta zero mode})
 \begin{equation}\label{eq:H I}
 	\mathscr{H}_I=\frac{a^4\lambda}{2\sqrt{V}}
	\sum_{\vec{k}} 
	\Delta \phi_0\,  \chi_{\vec{k}}\chi_{-\vec{k}},
 \end{equation}
where the interaction picture fields $\Delta\phi_0$ and $\chi_{\vec{k}}$ are now free fields, as in subsection \ref{sec:Free Fields}. Perturbative calculations are carried out by expanding $U_I$ to the desired order in the coupling constants. Say, to second order in $\lambda$
\begin{equation}
	U_I(t, -T)\approx \mathbb{1}
	-i\int_{-T}^t \mathscr{H}_I(t_1)\, dt_1
	-\int_{-T}^{t}dt_1 \int_{-T}^{t_1}dt_2\, \mathscr{H}_I(t_1) \mathscr{H}_I(t_2).
\end{equation}

Interactions also affect the vacuum state. We would actually like to calculate the expectation value of different operators in the vacuum state of the full interacting theory, rather than that of the free theory. In order to obtain the former from the latter, we shall use a well-known theorem  by Gell-Mann and Low \cite{GellMann:1951rw}: In the interaction picture we choose the initial state $|\psi(-T)\rangle$  to be the vacuum $|0\rangle$ of the free theory, allow the initial time $-T$ to approach $-\infty$, and switch on the interactions ``infinitely slowly"
 by multiplying the coupling constant $\lambda$ by $e^{\varepsilon\, t}$, where $\varepsilon$ is a positive parameter  that is taken to  zero at the end of the calculation.  The inclusion of this factor not only recovers the interacting vacuum from that of the free theory, but also regularizes some of the oscillatory integrals in the limit $T\to \infty$. For simplicity we shall not write down this factor explicitly in our integrals, and its presence shall remain   implicit.  

\section{Transition Amplitudes and Probabilities}
\label{sec:Transition Amplitudes}

One of the main focuses of particle physics is the $S$-matrix. The latter is the overlap between appropriately defined $in$ and $out$ particle states, but it can also be expressed as the matrix element of the time evolution operator in the interaction picture \cite{Weinberg:1995mt}, 
\begin{equation}
\langle \Phi_{out}|\Psi_{in}\rangle=\langle \phi | U_I(+\infty,-\infty)|\psi\rangle,
\end{equation}
where $|\phi\rangle$ and $|\psi\rangle$ are free particle states whose quantum numbers match those of the $out$ and $in$ states.  Whenever  such such a matrix element is non-zero for  single-particle state $|\psi\rangle$ and a multi-particle state $|\phi\rangle$, the particle described by $|\psi\rangle$ is unstable.

In an inflating spacetime there is no  static  $out$ region, because the time derivatives of the  scale factor (\ref{eq:a})  diverge  in the asymptotic future $t\to 0^-$. Therefore, it does not seem possible to define appropriate $out$ states, and there is no full analogue of an $S$-matrix. Nevertheless, the time evolution operator in the interaction picture between the asymptotic regions $t\to -\infty$ and $t\to 0^-$ is still well-defined, and one may compute its matrix elements between free states as well.  If the transition amplitude between state of the inflaton and any other state is non-zero, that would be an indication that the inflaton is unstable. 

To make this idea more precise, consider the vacuum expectation value of an arbitrary (hermitian) observable $\mathcal{O}$ in the presence of interactions,
\begin{equation}\label{eq:O vev I}
\langle \mathcal{O}(t)\rangle=\langle 0|U_I^\dag(t,-\infty)\mathcal{O}_I(t) U_I(t,-\infty)|0\rangle,
\end{equation} 
where $U_I$ is the time-evolution operator (\ref{eq:UI}), and the interaction picture operator $\mathcal{O}_I$ follows the free time evolution. There is no need to divide the expectation value by the $in$-$in$ amplitude $\langle 0 |U_I^\dag U_I|0\rangle$ because the latter equals one, as $U_I$ is unitary. For the same reason, in an expansion of the expectation value in terms of Feynman diagrams it suffices to consider connected diagrams alone.

It is convenient to expand the time-evolution operator $U_I$ into the identity plus a piece related to the interactions. Defining the operator $T$ by the relation 
$
U_I(t,-\infty)=\mathbb{1}+i T,
$
and inserting appropriate resolutions of the identity, the expectation value (\ref{eq:O vev I}) becomes
\begin{equation}\label{eq:vev}
	\langle \mathcal{O}(t)\rangle=
	\langle 0|\mathcal{O}_I|0\rangle
	- 2\,\mathrm{Im}\sum_\psi   \langle 0|\mathcal{O}_I|\psi\rangle   T_\psi 
	+\sum_{\psi,\psi'}
	T^*_{\psi'}\langle\psi'|\mathcal{O}_I |\psi\rangle T_\psi,
\end{equation}
where we have introduced the transition amplitude between the vacuum and a free state $|\psi\rangle$
\begin{equation}\label{eq:T}
	T_\psi\equiv \langle \psi | T |0\rangle.
\end{equation}
Equation (\ref{eq:vev}) shows that all we need to know to calculate the expectation value of any operator are its matrix elements in the free theory, $\langle\psi'|\mathcal{O}_I |\psi\rangle $, and the transition amplitudes to those states, $T_\psi$. If the transition amplitude $T_\psi$ is non-zero, it is as if the inflaton state had made a transition from $|0\rangle$ to $|\psi\rangle$, as expected from a decay.

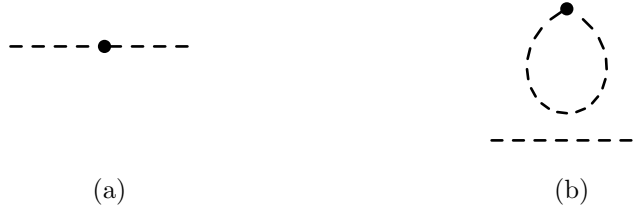
\begin{figure}
\subfigure[]
{
\begin{fmfgraph}(25,25) 
\fmfleft{v1} 
\fmfright{v2}
\fmf{dashes}{v1,vm}
\fmf{dashes}{vm,v2}
\fmfdot{vm}
\fmfforce{.5w,0.5h}{vm}
\end{fmfgraph}
}
\hspace{3cm}
\subfigure[] 
{
\begin{fmfgraph}(25,25) 
\fmfleft{v1,v1p} 
\fmfright{v2,v2p}
\fmf{phantom}{v1p,vm,v2p}
\fmfforce{.5w,0.7h}{vm}
\fmf{dashes}{v1,v2}
\fmfdot{vm}
\fmf{dashes,right,tension=0.6}{vm,vm}
\end{fmfgraph}
}
\caption{The two diagrams that represent the matrix element $\langle L|\phi_0^2|L\rangle$, where $|L\rangle=b^\dag|0\rangle$ is the state with a single zero mode inflaton quantum.  A dashed line stands for the inflaton zero mode, and a dot for the field insertion $\phi_0^2$.  Diagram (a) is clearly connected. The disconnected part of the expectation, $\langle L|\phi_0^2|L\rangle_\mathrm{disc}$, is the contribution of diagram (b). }\label{fig:disconnected} 
\end{figure}

Actually, some of the summands in equation (\ref{eq:vev}) cancel out and need not be considered. Because $U_I$ is unitary, $\langle U_I^\dag U_I\rangle=1$. Inserting in the last equation $U_I=1+iT$ yields  the ``optical theorem"
\begin{equation}
	2\,\mathrm{Im}\, T_0=\sum_{\psi} P_\psi\equiv P_\mathrm{tot},
\end{equation}
where $T_0$ is the vacuum persistence amplitude, 
$	P_\psi=|T_\psi|^2
$
the transition probability to the state $\psi$, and $P_\mathrm{tot}$ the total transition probability.   Because of  the optical theorem, the summand $-2\mathrm{Im}\, \langle 0 | \mathcal{O}_I|\psi\rangle T_0$  in equation (\ref{eq:vev}) is cancelled by the disconnected piece of  $\sum_{\psi,\psi'}
	T^*_{\psi'}\langle\psi'|\mathcal{O}_I |\psi\rangle T_\psi$. The  disconnected piece of a matrix element is defined by the relation
\begin{equation}\label{eq:disc}	
	\langle\psi_2|\mathcal{O}_I |\psi_1\rangle_\mathrm{disc}\equiv 
	\langle 0|\mathcal{O}_I |0\rangle \langle \psi_2 |\psi_1\rangle= 
	\langle 0|\mathcal{O}_I |0\rangle \delta_{\psi_2 \psi_1}.
\end{equation}
If we represent the matrix element  $\langle\psi_2|\mathcal{O}_I |\psi_1\rangle $ diagrammatically, its disconnected piece  is the contribution from disconnected diagrams, for which the external lines that represent the  states $\psi_1$ and $\psi_2$ simply go through the diagram, and are hence disconnected from the operator insertion $\mathcal{O}_I$ (see the example on figure \ref{fig:disconnected}.)  This cancellation is essentially the same  that allows disconnected diagrams to be disregarded in the $in$-$in$ formalism. In field theories in Minkowski spacetime the instability of a particle is also signaled by the appearance of a non-zero imaginary  component of its forward scattering amplitude. Here, the quantum state of the inflaton is not a one-particle state, but a coherent state with an infinite number of quanta. By shifting the inflaton field, the quantum this state becomes the vacuum $|0\rangle$, whose stability we expect to be  quantified by the vacuum persistence amplitude $T_0$.

\unitlength = 1mm 

\begin{figure}
\subfigure[]
{
\begin{fmfgraph}(25,25) 
\fmfleft{v1} 
\fmfright{v2}
\fmf{plain}{v1,v1}
\fmf{plain}{v2,v2}
\fmf{dashes}{v1,v2}
\end{fmfgraph}
}
\hspace{3cm}
\subfigure[] 
{
\begin{fmfgraph}(25,25) 
\fmfleft{i}
\fmfright{o}
\fmf{phantom,tension=10}{i,v1}
\fmf{phantom,tension=10}{v2,o}
\fmf{plain,left,tension=0.4}{v1,v2,v1}
\fmf{dashes}{v1,v2}
\end{fmfgraph}
}
\caption{The two diagrams that contribute to the vacuum persistence amplitude $T_0$ at leading order. A dashed line represents the inflaton zero mode and a solid line a matter field. By cutting these diagrams vertically through the middle we can identify the particles the vacuum can decay into, namely, a single inflaton zero mode, or  a zero mode  plus two matter quanta (see also figure \ref{fig:decays}.) Unless otherwise noted, all of our Feynman diagrams stand for transition amplitudes (and not $in$-$in$ expectation values.) }\label{fig:T0} 
\end{figure}
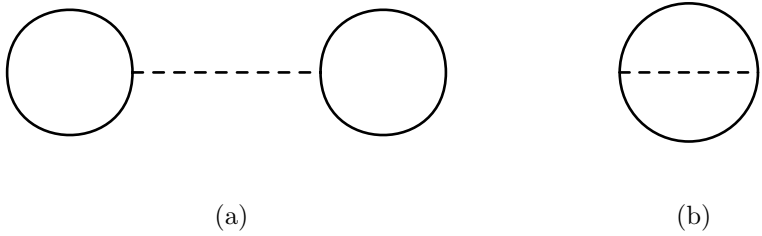

One of the main goals of this section is the calculation of  transition amplitudes and probabilities, not only because the former  capture  the intuitive concept of ``decay," but also because they automatically determine the expectation value of any observable. Two  of the observables we shall be  concerned about are the field deviation $\Delta\phi_0$ and its square  $\Delta \phi_0^2$. The former has a non-zero matrix element between the vacuum and a state with a single excitation of the inflaton zero mode, $|L\rangle=b^\dag |0\rangle$, so we shall be interested in the transition amplitude and probability  to such an excited state,
\begin{equation}\label{eq:T1}
 T_L\equiv \langle L|T|0\rangle, \quad P_L=|T_L|^2.
\end{equation}
Here, and in what follows, we shall refer to a single excitation of the inflaton's zero mode by ``$L$."  The matrix element  $\langle 0 |\Delta\phi_0^2|L\rangle$  vanishes. But since, $\Delta\phi_0^2$  has a non-zero matrix element with  states that contain two long quanta,   we shall also be interested in the transition amplitudes
\begin{equation}
T_{LL}\equiv \langle L, L|T|0\rangle\equiv
\frac{1}{\sqrt{2}}\langle 0 |b \, b\, T  |0\rangle, 
\end{equation}
as well as in the amplitudes and probabilities
\begin{equation}
	T_{L\vec{k}-\vec{k}}\equiv\langle L,  \vec{k} ,{}-\vec{k} |T|0\rangle, \quad P_{L\vec{k}-\vec{k}}=|T_{L\vec{k}-\vec{k}}|^2,
\end{equation}
where $|L, \vec{k}, -\vec{k}\rangle=b^\dag \,c^\dag_{\vec{k}}\, c^\dag_{-\vec{k}}|0\rangle$ is a state with a single inflaton zero mode and two matter quanta of opposite momenta $\vec{k}$.   The expectation value of $\Delta\phi_0^2$ in the state $|L, \vec{k},-\vec{k}\rangle$ does not depend on the value of $\vec{k}$, which is why we shall also encounter the  total decay probability into three quanta
 \begin{equation}\label{eq:P3}
	P_3=\frac{1}{2}\sum_{\vec{k}} P_{L\vec{k}-\vec{k}}. 
\end{equation}  
(We include a factor of $1/2$ in the sum because $|L,\vec{k}, -\vec{k}\rangle$ and $|L,-\vec{k}, \vec{k}\rangle$ represent the same state.) At lowest order in $\lambda$, the sum of $P_L$ and $P_3$  is simply the total decay probability of the inflaton, $P_\mathrm{tot}$. For a cubic interaction of the form (\ref{eq:H I}) the only two diagrams that contribute to $T_0$ at leading order ($\propto\lambda^2$) are those on figure \ref{fig:T0}. Cutting both diagrams vertically through the middle reveals the final states the inflaton can decay into, namely,  a single zero mode quantum or  a zero mode plus two matter quanta. These are of course the decays whose probability is captured by $P_L$ and $P_3$. Finally, we shall also need to calculate the transition amplitude into two matter fields $|\vec{k},-\vec{k}\rangle=c^\dag_{\vec{k}}\,c^\dag_{-\vec{k}}|0\rangle$,
\begin{equation}
T_{\vec{k}, -\vec{k}}\equiv \langle \vec{k},-\vec{k} |T|0\rangle,
\end{equation}
which is of order $\lambda^2$ and enters the leading order correction to the energy-momentum tensor of the inflaton decay products. In flat spacetime, all the transition amplitudes above would vanish because of energy conservation. In an expanding background energy is not conserved, so these transitions are allowed. Because  spatial translations remain isometries, though, spatial momentum is conserved, which is why the matter fields quanta appear in pairs of opposite momenta.  In some instance, relying on transition amplitudes  reduces and better organizes the number of diagrams needed to be considered, and is thus computationally simpler than a direct calculation of the expectation value in the $in$-$in$ formalism. 

\unitlength = 1mm 

\subsection{Decay at First order}

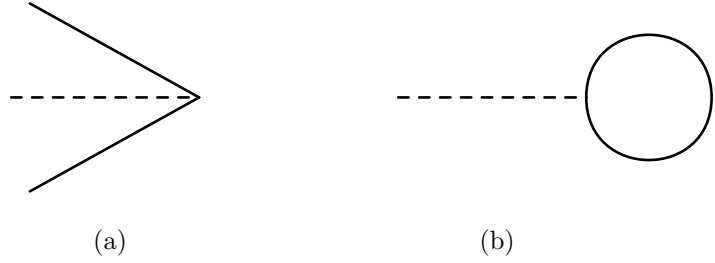
\begin{figure}
\subfigure[]
{
\begin{fmfgraph}(25,25) 
\fmfleft{k,L,mk} 
\fmfright{v}
\fmf{plain}{k,v}
\fmf{dashes}{L,v}
\fmf{plain}{mk,v}
\end{fmfgraph}
}
\hspace{2cm}
\subfigure[] 
{
\begin{fmfgraph}(25,25) 
\fmfleft{L}
\fmfright{v}
\fmf{dashes}{L,v}
\fmf{plain}{v,v}
\end{fmfgraph}
}
\caption{The two  decay channels at first order in $\lambda$. (a) Decay into two matter quanta plus an inflaton zero mode. (b) Decay into a single inflaton zero mode. To match the structure of the matrix elements of $T$, initial states appear on the right of the diagram, and final states on the left.}\label{fig:decays} 
\end{figure}

At first order in $\lambda$, the only two possible final states are $|L, \vec{k},-\vec{k}\rangle$ and $|L\rangle$, as shown in figure \ref{fig:decays}. To render the analytical calculations somewhat more manageable, we are going to consider two opposite limits: The limit in which the effective mass of the decay products is much larger than the Hubble scale $H$, and the limit in which the effective mass is much smaller than $H$. 
By equation (\ref{eq:meff}), the effective mass of matter particles is determined by the inflaton field, so for fixed $\lambda$, each limit can be regarded as a limit of large or small inflaton values or coupling constants.   Note that our results only depend on the effective mass of the particles $\chi$, and not on the origin of this mass. In the presence of a ``bare" mass term $m_0^2 \,\chi^2$ in the action of the theory (\ref{eq:S}), the effective mass becomes $m_\chi^2=m_0^2+\lambda \bar{\phi}$, and all our results carry through by using the last expression instead of  equation (\ref{eq:meff}).

We begin by evaluating the transition amplitude to the state $|L,\vec{k},-\vec{k}\rangle$ to lowest order,
\begin{equation}
	iT_{L\vec{k}-\vec{k}}\equiv -i \int\limits_{-\infty}^ t dt_1 \, \langle L, \vec{k},-\vec{k}| \mathscr{H}_I(t_1) |0\rangle
	=-i\frac{\lambda}{\sqrt{V}} \int\limits_{-\infty}^t dt_1 \,a^4(t_1) u^*(t_1)w_k^*{}^2(t_1),
\end{equation}
where we have used that the interaction Hamiltonian is that of equation (\ref{eq:H I}).  Substituting the form of the mode functions and changing integration variables to $z_1=-k t_1$ this becomes
\begin{equation}\label{eq:T inflation}
	T_{L\vec{k}-\vec{k}}=-\frac{1}{2\sqrt{2 k_L^3 V}}\frac{\lambda}{H} \int_{-kt}^{\infty} \frac{dz_1}{z_1}\, \left[1-\frac{i}{3}\left(\frac{k_L}{k}z_1\right)^{3-\eta/3} \right]v^*{}^2(z_1),
\end{equation}
which happens to depend on $\vec{k}$ only through the combination $kt$. We then  obtain the  decay probability into three quanta (\ref{eq:P3}) by adding all the individual probabilities, 
\begin{equation}\label{eq:P inflation}
 	P_3=\frac{1}{2}\sum_{\vec{k}} |T_{L\vec{k}-\vec{k}}|^2\approx \frac{V}{4\pi^2}\frac{1}{(-t)^3}\int_0^\infty dz\, z^2
|T_{L\vec{k}-\vec{k}}(z)|^2,
\end{equation}
where we have approximated the sum over $\vec{k}$ by an integral and, again, ${z=-kt}$. For large $z_1$, the function $v$ in equation (\ref{eq:T inflation}) approaches the flat spacetime limit $e^{iz_1}/\sqrt{z_1}$, which implies that at large $-kt$ the transition amplitude $T_{L\vec{k}-\vec{k}}(z)$ is of  order $e^{-2iz}/z^2$.  This renders the integral over $z$ in (\ref{eq:P inflation}) convergent.  

Things are quite different for the transition amplitude $T_L$ in equation (\ref{eq:T1}).  At lowest order in $\lambda$, the latter is instead
\begin{equation}\label{eq:T1ren}
	T_L= - 
	\int_{-\infty}^ t dt_1\, \langle L| \mathscr{H}_I(t_1) |0\rangle
	\approx
	-\frac{\lambda\sqrt{V}}{2H^4}\langle\chi_I^2\rangle\int_{-\infty}^t \frac{dt_1}{t_1}\frac{u^*(t_1)}{t_1^3},
\end{equation}
where we have used that the expectation of $\chi_I^2(t,\vec{x})$ in de Sitter is space and time independent,
\begin{equation}\label{eq:chi sq}
	\langle \chi_I^2(t,\vec{x})\rangle=\frac{H^2}{4\pi^2}\int_0^\infty dz\, z^2 |v(z)|^2.
\end{equation}
The  proportionality of $T_L$ to $\langle\chi_I^2\rangle$ at this order can be seen on diagram (b) in figure \ref{fig:decays}. 
Some care must be taken in the evaluation of the time integral in (\ref{eq:T1ren}), because we shall later need its imaginary part, which is proportional to the decaying mode, whose integral diverges in the strict massless limit $\eta=0$.  Keeping track of the decaying mode we arrive at 
\begin{equation}\label{eq:uc int}
	\int_{-\infty}^t
	\frac{dt_1}{t_1}\frac{u^*(t_1)}{t^3_1}
	\approx
	\frac{H \, k_L^{3/2}}{\sqrt{2}}
	\left[\frac{1}{3}\frac{1}{(-k_L t)^{3}}
	-\frac{i}{\eta}\frac{1}{(-k_Lt)^{\eta/3}}\right].
\end{equation}
To calculate $\langle \chi_I^2\rangle$ in equation (\ref{eq:chi sq}), we note that for large $z$,  $|v(z)|^2$ approaches $1/z$, and the integral over $z$ diverges quadratically. To make sense of $\langle \chi_I^2\rangle$  (and thus $T_L$) we  need to regularize and renormalize. As we mentioned in the introduction, in this work we  rely on adiabatic subtraction \cite{Parker:1974qw,Birrell:1982ix}, which takes care of both steps at once. In this approach, we subtract from the transition amplitude  the expression obtained by replacing the mode functions by adiabatic approximations. The adiabatic order of the approximations is simply set by the requirement that the subtracted expression be finite for any of the free parameters of the theory. In the case at hand, it thus suffices to subtract the second adiabatic order  approximation $v^{(2)}$,
\begin{equation}\label{eq:chiren}
	\langle \chi_I^2\rangle^\mathrm{ren}=\frac{H^2}{4\pi^2}\int_0^\infty z^2 \left(|v(z)|^2-|v^{(2)}(z)|^2\right).
\end{equation}
where the form of the adiabatic modes $v^{(2)}$ follows from the results in Appendix \ref{sec:Adiabatic Modes},
\begin{equation}\label{eq:v2ad}
	|v^{(2)}(-kt)|^2= -\frac{1}{t}\left[\frac{1}{\omega_0}-\frac{1}{2\omega_0^3}\left(\frac{3}{4}\frac{\dot{\omega}_0^2}{\omega_0^2}-\frac{1}{2}\frac{\ddot{\omega}_0}{\omega_0}-\frac{\ddot{a}}{a}\right)\right].
\end{equation}
To emphasize that a given quantity has been renormalized in the adiabatic scheme, we append the superscript ``ren" to it. Combining  equations (\ref{eq:T1ren}) and (\ref{eq:uc int}) we thus get a relation between $T_L$ and the renormalized value of $\langle \chi_I^2\rangle$,
\begin{equation}\label{eq:TL}
T_L^\mathrm{ren}=-\frac{1}{6}\sqrt{\frac{k_L^3 V}{2}}
	\frac{\lambda\langle\chi_I^2\rangle^\mathrm{ren}}{H^3}\frac{1}{(-k_L t)^3}\left[1
	-\frac{3i}{\eta}(-k_Lt)^{3-\eta/3}\right].
\end{equation}  
Note the divergence with $1/\eta$, which is why we had to avoid the limit $\eta\to 0$ in the decaying mode of $u$. 

\subsubsection{Limit of Heavy Fields}

We proceed now to the evaluation of the different transition amplitudes and probabilities in the  limit of heavy matter fields, $m_\chi\gg H$. In this limit it is useful to approximate the Hankel function by the uniform expansion in Appendix \ref{sec:Uniform Expansion}. In the heavy field limit, it  suffices to keep just the first term in the expansion
\begin{equation}\label{eq:Dunster}
	v(z)\approx 
	\frac{e^{i\,\mu\, \xi(z)}}{(\mu^2+z^2)^{1/4}}+\mathcal{O}(\mu^{-2}).
\end{equation}
The  integral over $z_1$  in equation (\ref{eq:T inflation}) is then highly oscillatory, but there are no points where the phase $\xi(z)$ is stationary. We can evaluate the integral instead by repeated integration by parts, which also results in an asymptotic expansion in powers of the small parameter $\mu^{-1}$,
\begin{equation}
\int\limits_{-\infty}^z \,  dz_1 \,f(z_1)\, e^{-2i\mu \xi(z_1)}=\left[\frac{f(z_1)}{-2i\mu \,d\xi/dz_1} e^{-2i\mu \xi(z_1)}\right]^z_{-\infty}
-\int\limits_{-\infty}^z \,  dz_1 \frac{d}{dz_1}\left(\frac{f(z_1)}{-2i\mu\, d\xi/dz_1}\right)e^{-2i\mu \xi(z_1)}.
\end{equation}
 At next to lowest order in $\mu^{-1}$ we find
\begin{equation}\label{eq:Tk heavy}
T_{L\vec{k}-\vec{k}}(z)\approx 
	\frac{i}{4 \sqrt{2k_L^3 V}}\frac{\lambda}{H}
	\frac{e^{-2i\mu \xi(z)}}{\mu^2+z^2}
	\left[\left(1-\frac{i}{3}(-k_L t)^{3-\eta/3}\right)
	+\frac{i z^2}{(\mu^2+z^2)^{3/2}}+\cdots\right], 
\end{equation}
which reaches its largest magnitude in the long-wavelength limit $z=-kt\to 0$, and is suppressed by the large ratio $\mu^2\approx m_\chi^2/H^2$.  Inserting this amplitude into the decay probability  (\ref{eq:P inflation}) results in 
\begin{equation}\label{eq:P3 heavy}
	P_3\approx\frac{1}{512\pi}\frac{\lambda^2}{m_\chi \,H} \frac{1}{(-k_L t)^3}\left[1+\frac{9}{\eta^2}(-k_L t)^{6-2\eta/3}\right],
\end{equation}
where we only quote the leading terms.  

  Although the probability is suppressed by the small factor $\lambda^2 /m_\chi H$,  this suppression factor is more than compensated by   $(- k_L t)^3\equiv e^{3N_L}$. The latter is simply the exponential of the number of e-folds since a mode of the size of the entire (finite) universe left the horizon. Since our finite universe must encompass the visible universe, $N_L$ must be larger than about fifty.  We thus conclude that for reasonable parameter choices, the decay probability should be exponentially large.  At this point the reader may recall Weinberg's work on the future asymptotic behavior of quantum correlators during inflation \cite{Weinberg:2005vy,Weinberg:2006ac}.  He argued that as long as field interactions are not proportional to too many powers of the scale factor $a$, quantum corrections to $in$-$in$ correlators cannot become large. Our transition amplitude is not an expectation value yet, but in any case, as also noted in \cite{Weinberg:2005vy}, a non-derivative  interaction of the form $\sqrt{-g}\, \lambda\phi\chi^2$ does not satisfy Weinberg's conditions of convergence. A possible way to avoid the large enhancement of the decay probability could involve derivative couplings of the inflaton to matter, such as $\sqrt{-g}\,  \phi \partial_\mu \chi \partial^\mu \chi$. Since the latter is proportional to $a^2$, rather than $a^4$, this is likely to tame the exponential growth of the decay probability. 
As already pointed out in \cite{Weinberg:2005vy} such derivative couplings are the only possible ones if $\chi$ is a Goldstone boson.   

Let us turn our attention now to the renormalized transition amplitude (\ref{eq:T1ren}), which we obtain by replacing $\langle \chi^2_I\rangle$ by its renormalized counterpart.  Using equation  (\ref{eq:v2ad}) we find that
\begin{equation}\label{eq:v2adsq}
 |v^{(2)}(z)|^2= \frac{1}{\sqrt{m_H^2+z^2}}\left(1+\frac{9 m_H^4+22 m_H^2 z^2+8 z^4}{8 (m_H^2+z^2)^3}\right),
\end{equation}
where $m_H\equiv m_\chi/H$.  Then, from  the asymptotic expansion (\ref{eq:Dunster}) with $n=4$ and the second order adiabatic modes (\ref{eq:v2adsq})  we obtain  by brute-force calculation that in the limit of large $m_\chi$,
\begin{equation}\label{eq:int diff heavy}
	\langle \chi^2(t,\vec{x})\rangle^\mathrm{ren}
	\approx \frac{29}{60} \frac{H^2}{m_\chi^2}
\left(\frac{H}{2\pi}\right)^{2}.
\end{equation}
Cancellations among the different terms in $|v^{(2)}(z)|^2$ and $|v(z)|^2$ yield an integral of order $1/\mu^2$, which is why we had to keep terms in the uniform expansion  to order $\mu^{-4}$. 

To conclude, let us revisit the decay probability $P_3$. Later on we shall need to calculate the expectation of an observable that depends on $P_3$. The former turns out to contain an additional divergent integral that needs to be renormalized by subtraction of the zeroth order adiabatic modes. Therefore, because in the adiabatic scheme one subtracts from the divergent expectation value its adiabatic approximation,  in that context we should  subtract from $P_3$ the zeroth order adiabatic approximation too. Since the zeroth-order adiabatic modes in the heavy field limit have the same functional form as the exact modes, with $\mu$ simply replaced by $m_\chi/H$,  it is easier instead to replace  the integrand  by its derivative with respect to $\mu$ times $\mu-m_\chi/H\sim H/m_\chi$ . The derivative lowers the degree of divergence of the expression, which in the case at hand becomes
\begin{equation}
	P_3^\mathrm{ren}=\frac{9}{4096\pi}\frac{\lambda^2}{H^2}\frac{H^3}{m_\chi^3}
	\frac{1}{(-k_L t)^3}
	\left[1+\frac{9}{\eta^2}(-k_L t)^{6-2\eta/3}\right].
\end{equation}
Note the suppression by an additional factor of $H^2/m_\chi^2$ as compared to (\ref{eq:P3 heavy}), as could have been guessed from the the subtraction procedure we just described.

\subsubsection{Limit of Light Fields}

In the limit of light matter fields,  $m\chi\ll H$, the mode functions are 
\begin{equation}\label{eq:v light}
	v(z)\approx \frac{e^{iz}}{\sqrt{z}}\left(1+\frac{i}{z}\right).
\end{equation}
Therefore, in the long-wavelength limit $-kt\ll 1$ the transition amplitude  in equation (\ref{eq:T inflation}) readily evaluates to
\begin{equation}\label{eq:Tk light}
T_{L\vec{k}-\vec{k}}=\frac{1}{6\sqrt{2k_L^3V}}\frac{\lambda}{H}\frac{1}{(-kt)^3}\left[1-\frac{3i}{\eta}(-k_L t)^{3-\eta/3}+\cdots\right].
\end{equation}
In this case the amplitude is strongly enhanced in the long wavelength limit, by  the characteristic factor $1/(-k t)^3$. 

The different behavior of the transition amplitude in the long wavelength limit has drastic implications, namely,  the total decay probability into a pair of matter quanta (\ref{eq:P inflation}) blows up in the infrared. Infrared divergences are typical of massless theories, but in this case the divergence survives away from the massless limit $m_\chi=0$. For a finite mass, at small $z$, 
$v(z)\sim z^{-\nu}$ and $T_{L\vec{k}-\vec{k}}\sim (-kt)^{-2\nu}$.
Therefore, although the  integral (\ref{eq:P inflation}) converges  in the ultraviolet $z\to \infty$, where it has the same behavior as in the heavy field limit, it diverges in the infrared,  whenever $\nu>3/4$.  Since we are dealing with  a finite universe here, the infrared divergence is  just an artifact of our continuum approximation.  In particular, in a finite universe of size $L$, the smallest (non-zero) value of $k$ is $k_{IR}=2\pi/L$. Imposing an infrared cut-off at $k_{IR}$, and  focusing in the dominant contribution in the infrared, we obtain instead
\begin{equation}\label{eq:P3 light}
	P_3\approx\frac{1}{864\pi^2}\frac{\lambda^2}{H^2}
 \frac{1}{(-k^3_L t)^3 (-k_{IR} t)^3}\left[1+\frac{9}{\eta^2}(-k_L t)^{6-2\eta/3}\right],
\end{equation}
where we have returned to the limit $m_\chi\to 0$. Incidentally, with $k_{IR}=2\pi/L$, equation (\ref{eq:P3 light}) is basically what we would have gotten by including only the longest modes in the discrete sum of equation (\ref{eq:P3}).  We keep $k_L$ and $k_{IR}$ as separate quantities to convey how our results depend on the normalization of the zero mode and the value of the infrared cut-off, but note that both are expected to be of order $1/L$. In that case, the probability grows   as  $e^{6N_L}$, where $N_L$ is the number of e-folds of inflation since the universe left the horizon. Clearly, such an   exponential enhancement is likely to overcome any eventual suppression of the probability by $\lambda$. 

We turn our attention now to the amplitude for a transition between the vacuum and a single excitation of the inflaton zero mode, equation (\ref{eq:T1ren}). It can be readily checked that  for massless fields the difference between the exact mode functions and their second order adiabatic approximation vanishes, thus implying $\langle \chi^2_I\rangle^\mathrm{ren}=0$.  Away from the strict massless limit, the integral (\ref{eq:chiren}) remains finite, as can be checked by inspecting the integrand in the ultraviolet ($z\to\infty$) and infrared ($z\to 0$) limits.  The integral is dominated by the exact modes $z^2|v|^2\sim z^{2-2\nu}$, whose contributions diverges as $z\to 0$, while $z^2 |v^{(2)}|^2$ approaches zero. Hence, in the limit $\nu\to 3/2$, provided that $k_{IR}\ll m_\chi$, the integral over $z$ is of order 
\begin{equation}\label{eq:int diff light}
	\langle \chi^2(t,\vec{x})\rangle^\mathrm{ren}
\approx \frac{3}{2} \frac{H^2}{m_\chi^2} \left(\frac{H}{2\pi}\right)^{2}.
\end{equation}
Although this expression appears to blow up in the limit $m_\chi \to 0$, this is just an artifact of the infinite volume limit $k_{IR}=0$, which is necessary for $k_{IR}\ll m_\chi$ to hold for all masses $m_\chi$. In a finite volume universe, as $m_\chi \to 0$ the difference between the exact and adiabatic modes  approaches zero at all values of $k$. In such a way, $\langle \chi^2\rangle^\mathrm{ren}$ remains continuous at $m_\chi=0$.
 
It is  in fact reassuring that we can derive some of the previous results using a somewhat different method.  The calculation of the expectation value of $\langle \chi_I^2(t,\vec{x})\rangle$ amounts to the calculation of the propagator of $\chi$ in the limit of coincident points. The latter diverges, with  coefficients proportional to different curvature invariants. To eliminate these divergences, one subtracts from the coincident limit an appropriate ``adiabatic" short-distance expansion of the propagator.   Typically one is interested in calculating the renormalized action,  or the renormalized energy-momentum tensor, and one needs to subtract an  adiabatic expansion of the propagator to fourth order. The subtraction leaves a finite result that can be taken to be the renormalized value of the expectation value.  Here, since we are just interested in $\langle \chi_I^2\rangle$ by itself, it suffices to subtract an expansion to second adiabatic order.\footnote{In other words, the counterterms we would need to renormalize $\langle \partial \mathscr{H}_I/\partial \phi_0\rangle\propto\langle \chi^2\rangle $ are not the same as those needed to renormalize $\langle T_{\mu\nu}\rangle \supset \sqrt{-g}\langle \chi^2\rangle$. The former would involve curvature invariants proportional to $\phi$, whereas the latter would involve  curvature invariants alone.}
In de Sitter spacetime, the renormalized value of $\langle \chi_I^2\rangle$ calculated as described is  (see equation (6.182) in \cite{Birrell:1982ix})
\begin{equation}\label{eq:chisq vev}
	\langle \chi_I^2(t,\vec{x}) \rangle^\mathrm{ren}=
	\frac{H^2}{16\pi^2}\left\{\left(\frac{m_\chi^2}{H^2}-2\right)\left[\psi\left(\frac{3}{2}+\nu\right)+\psi\left(\frac{3}{2}-\nu\right)-\log \frac{m_\chi^2}{H^2} -1\right]+\frac{m_\chi^2}{H^2}-\frac{2}{3}\right\},
\end{equation}
where $\psi\equiv \Gamma'/\Gamma$ is the digamma function and we have restored the (finite)  term of fourth adiabatic order.  In the limit of light and heavy fields this  
 agrees with  equations (\ref{eq:int diff heavy}) and (\ref{eq:int diff light}).

\subsection{Decay at Second Order}
\begin{figure}
\subfigure[]
{
\begin{fmfgraph}(25,25) 
\fmfleft{k,mk} 
\fmfright{v1,v2}
\fmf{plain}{k,v1}
\fmf{plain}{mk,v2}
\fmf{plain,right}{v1,v2}
\fmf{dashes}{v1,v2}
\end{fmfgraph}
}
\hspace{1cm}
\subfigure[]
{
\begin{fmfgraph}(25,25) 
\fmfleft{k,mk} 
\fmfright{v2}
\fmf{plain}{k,v1}
\fmf{plain}{mk,v1}
\fmf{dashes}{v1,v2}
\fmf{plain}{v2,v2}
\end{fmfgraph}
}
\hspace{1cm}
\subfigure[] 
{
\begin{fmfgraph}(25,25) 
\fmfleft{k,mk} 
\fmfright{v1,v2}
\fmf{dashes}{k,v1}
\fmf{dashes}{mk,v2}
\fmf{plain}{v1,v1}
\fmf{plain}{v2,v2}
\end{fmfgraph}
}
\hspace{1cm}
\subfigure[] 
{
\begin{fmfgraph}(25,25) 
\fmfleft{k,mk} 
\fmfright{v1,v2}
\fmf{dashes}{k,v1}
\fmf{dashes}{mk,v2}
\fmf{plain,right}{v1,v2}
\fmf{plain}{v1,v2}
\end{fmfgraph}
}
\caption{The two  relevant decay channels at second order in $\lambda$. (a) and (b): Decay into two matter quanta. (c) and (d): Decay into a pair of inflaton zero mode quanta.}\label{fig:decays second} 
\end{figure}
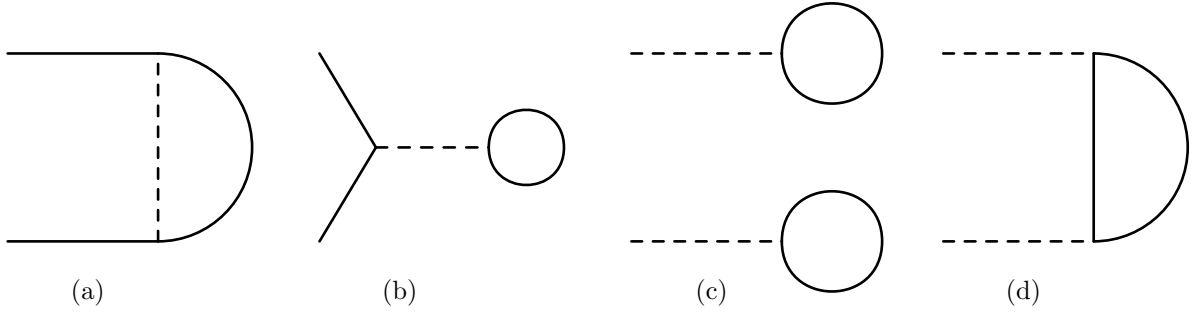

At second order in $\lambda$, the vacuum can decay into a pair of quanta, as shown  in figure \ref{fig:decays second}. The transition amplitude to a pair of matter quanta is the sum of diagrams (a) and (b) in figure \ref{fig:decays second},
\begin{equation}
	T_{\vec{k}-\vec{k}}={}^{(a)}T_{\vec{k}-\vec{k}}+{}^{(b)}T_{\vec{k}-\vec{k}}.
\end{equation}
Looking at diagram (a), or directly from the corresponding expressions for the transition amplitude we find 
\begin{equation}\label{eq:Tkk}
 	{}^{(a)}T_{\vec{k}-\vec{k}}=\frac{-i}{\sqrt{2k_L^3 V}} \frac{\lambda}{H}\int_{-kt}^\infty \frac{dz_1}{z_1} 
	\left[1+\frac{i}{3}\left(-\frac{k_L z}{k}\right)^{3-\eta/3}\right]
	|v(z_1)|^2 
	\, T_{L\vec{k}-\vec{k}}(z_1).
\end{equation} 
The integral in (\ref{eq:Tkk}) can be readily evaluated using our previous methods. Since there is no mode sum, no renormalization is required. In the limit of heavy fields, the subleading corrections in the limit $-k_L t\ll 1$ cancel, while they survive in the light field limit, in which we only quote the dominant terms when $-k t\ll1$,
\begin{equation}
	 {}^{(a)}T_{\vec{k}-\vec{k}}=\frac{-i}{k_L^3 V}\frac{\lambda^2}{H^2}\times
	 \begin{dcases}
	 \frac{1}{16}\frac{e^{-2i\mu \xi(-kt)}}{(\mu^2+k^2t^2)^2}& \text{if } H\ll m_\chi, \\
	 \frac{1}{72}\frac{1}{(kt)^6}\left[1-\frac{6i}{\eta}(-k_L t)^{3-\eta/3}\right] & \text{if } m_\chi \ll H.
	 \end{dcases}
\end{equation}
The contribution from diagram (b) can be evaluated along the same lines. Since there is a mode sum from the closed matter field loop, the latter needs to be renormalized, which is why diagram (b) is proportional to $\langle \chi_I^2\rangle^\mathrm{ren}$, 
\begin{equation}
	{}^{(b)}T^\mathrm{ren}_{\vec{k}-\vec{k}}=
	\frac{\lambda^2 \langle \chi_I^2\rangle^\mathrm{ren}}{H^4}
	\frac{1}{(-k_L t)^3}
	\begin{dcases}
	\frac{1}{48}
	\frac{e^{-2i\mu\xi}}{\mu^2+k^2t^2}\left[1-\frac{3i}{\eta}(-k_L t)^{3-\eta/3}\right]
	& \text{if } H\ll m_\chi, \\
	\frac{-i}{72}\frac{1}{(-k t)^3}\left[1-\frac{6i}{\eta}(-k_L t)^{3-\eta/3}\right] &  \text{if } m_\chi\ll H.
	\end{dcases}
\end{equation}
Note that in the limit of constant $u$, this amplitude obeys ${}^{(b)}T^\mathrm{ren}_{\vec{k}-\vec{k}}=iT_{L\vec{k}-\vec{k}} T^\mathrm{ren}_L$, where $T_{L\vec{k}-\vec{k}}$ and $T^\mathrm{ren}_L$ are the first-order transition amplitudes in equations (\ref{eq:T inflation}) and (\ref{eq:TL}). Such a relation  could have been guessed from the structure of  diagram (b). 

The transition amplitude to two zero mode quanta $T_{LL}$ is the sum of the contributions from diagrams (c) and (d) in figure \ref{fig:decays second}, 
\begin{equation}
	T_{LL}={}^{(c)}T_{LL}+{}^{(d)}T_{LL}.
\end{equation}
It is relatively straightforward to  evaluate  the renormalized contribution to $T_{LL}$ from diagram (c) for any matter field mass,
\begin{equation}\label{eq:TLLb}
	{}^{(c)}T^\mathrm{ren}_{LL}=
	\frac{i}{\sqrt{2}} \frac{k_L^3 V}{72}  
	\frac{\lambda^2 \langle\chi_I^2\rangle^2_\mathrm{ren}}{H^6}
	\frac{1}{(k_L t)^6}
	\left[1-\frac{3i}{\eta}(-k_L t)^{3-\eta/3}\right].
\end{equation} 
Neglecting the decaying mode in $u$ amounts to keeping only the leading term in the previous expression, the one proportional to $(k_L t)^{-6}$. In this approximation, it is readily seen that ${}^{(c)}T^\mathrm{ren}_{LL}=\frac{i}{\sqrt{2}}(T_L^\mathrm{ren})^2$,
where $T^\mathrm{ren}_L$ is the (renormalized) transition amplitude into an inflaton zero mode to first order in $\lambda$, which we have already calculated in the previous subsection. This is in fact what diagram (c) appears to suggest.   The second contribution,  $^{(d)}T_{LL}$, cannot be that easily recovered from previous amplitudes, and needs to be evaluated explicitly,
\begin{equation}\label{eq:dTLL}
	{}^{(d)}T_{LL}=\frac{-i}{\sqrt{2}}\frac{\lambda}{\sqrt{V}} \int^t dt_1\, a^4(t_1) u^*(t_1) \sum_{\vec{k}} w_k^2(t_1)T_{L\vec{k}-\vec{k}}(t_1).
\end{equation}
If $u$ were constant, the integrals in equation (\ref{eq:dTLL}) would converge, but because of the decaying mode they do not, and it is necessary to renormalize by subtraction of the  zeroth order adiabatic modes. As before, the subtraction does not have much of an effect on the dominant terms in the limit of light fields, which are dominated by the infrared cut-off,
\begin{equation}
	{}^{(d)}T^\mathrm{ren}_{LL}\approx
	\frac{(-k_L t)^{-3}}{\sqrt{2}\pi^2}
	\frac{\lambda^2}{H^2}\times
	\begin{dcases}
	\frac{3}{256}\frac{H}{m_\chi}\left[\left(1-\frac{6i}{\eta}(-k_L t)^{3-\eta/3}\right)
	+\frac{3\pi i}{16}\frac{H^2}{m_\chi^2}+\cdots\right]
	& \text{if } H\ll m_\chi,\\
	\frac{i}{864}\frac{1}{(-k_{IR} t)^3} \left(1-\frac{6i}{\eta}\, (-k_L t)^{3-\eta/3}\right)+\cdots
	& \text{if }m_\chi\ll H.
	\end{dcases}
\end{equation} 
Note that we have kept the subleading term in the heavy field limit  because $^{(d)}T^\mathrm{ren}_{LL}$ will later appear in combination with  $P_3^\mathrm{ren}$, which is of order $\mu^{-3}$.

\section{Expectation Values}

We have framed  our analysis so far in terms of decays of the inflaton into states with definite number of quanta. The concept of particle states plays a central role in $S$-matrix theory, but it faces its limits in curved spacetimes, due to the global nature of the particle concept \cite{Birrell:1982ix}. In addition, as we have argued, in an inflationary spacetime there is no static $out$ region, so it remains unclear what to make of the exponentially growing decay probabilities that we have encountered.  

But in any case, given that inflation is formulated purely as a field theory, it is questionable whether particles should play any role in its description. In the end, all we are interested in is expectation values of different field operators, which is all we need to cast the predictions of the theory.  This is the focus of the present section. 

\subsection{Zero mode}

In the previous subsection we have seen that the probability for the inflaton zero mode to decay is sizable. In order to study the impact of these transitions on the zero mode itself, we shall calculate the expectation of $\Delta\phi_V=\Delta\phi_0/\sqrt{V}$ and $\Delta\phi_V^2=\Delta\phi_0^2/V$. The first captures how these transitions affect the mean,  ``classical", value of the inflaton, whereas the latter tells us to what extent the field itself behaves classically. Of course, we could have derived both expectations directly from equation (\ref{eq:O vev I}). 

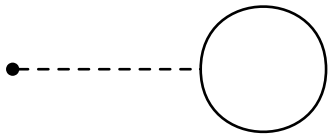
\begin{figure}
\begin{fmfgraph}(25,25) 
\fmfleft{v1} 
\fmfright{v2}
\fmf{plain}{v2,v2}
\fmf{dashes}{v1,v2}
\fmfdot{v1}
\end{fmfgraph}
\caption{Diagrammatic representation of the first order correction to the expectation of $\Delta\phi_0$ in the $in$-$in$ formalism. The dot denotes the insertion of $\Delta\phi_0$. This tadpole diagram is essentially diagram (a) of figure \ref{fig:T0} cut in half.}\label{fig:vev} 
\end{figure}

We begin by noting that in the free theory $\langle 0|\Delta\phi_0|\psi\rangle$ is nonzero only if $|\psi\rangle$ describes a single excitation of the inflaton zero mode (this is represented diagrammatically in figure \ref{fig:vev}.) Therefore, using equation (\ref{eq:vev}) we immediately infer that to leading order in the interaction
\begin{equation}\label{eq:delta phi vev}
	\langle \Delta\phi\rangle=-2\, \frac{\mathrm{Im}\,(u\, T_L)}{\sqrt{V}},
\end{equation}
where $T_L$ is the (renormalized) decay amplitude into a single zero mode excitation in  equation (\ref{eq:TL}). Inserting the latter into (\ref{eq:delta phi vev}) we  find, both in the limit of heavy and light fields  that
\begin{equation}\label{eq:delta phi}
	\langle \Delta\phi\rangle^\mathrm{ren}\approx
	-\frac{\lambda}{2\eta}\frac{\langle\chi_I^2\rangle^\mathrm{ren}}{H^2}
		\frac{1}{(-k_L t)^{\eta/3}}.
\end{equation}
Since $T_L$ grows as $e^{3N_L}$, one may have naively expected $\langle \Delta\phi\rangle$ to grow similarly, since the non-decaying mode of $u$ is constant. Instead, the leading (real) term in the $-k_L t\ll 1$ limit drops out, and we are left with a secular growth proportional to $e^{N_L\eta/3}$ stemming from the (imaginary) decaying mode of $u$. Thus, rather than growing with a power of the scale factor, as the transition amplitudes, in the limit  $N_L\eta\ll 1$ the expectation value grows with the logarithm of $a$. We shall  further analyze the implications of (\ref{eq:delta phi}) in the conclusions.  In the meantime,  note that the next to leading correction to $\langle \Delta\phi \rangle$ is of order $\lambda^3$.

\begin{figure}
\subfigure[]
{
\begin{fmfgraph}(25,25) 
\fmfleft{v1} 
\fmfright{v2}
\fmf{plain}{v1,v1}
\fmf{plain}{v2,v2}
\fmf{dashes}{v1,vm}
\fmfdot{vm}
\fmf{dashes}{vm,v2}
\end{fmfgraph}
}
\hspace{2cm}
\subfigure[] 
{
\begin{fmfgraph}(25,25) 
\fmfleft{i}
\fmfright{o}
\fmf{phantom,tension=10}{i,v1}
\fmf{phantom,tension=10}{v2,o}
\fmf{plain,left,tension=0.4}{v1,v2,v1}
\fmf{dashes}{v1,vm}
\fmfdot{vm}
\fmf{dashes}{vm,v2}
\end{fmfgraph}
}
\caption{The two diagrams that contribute to $\langle\Delta\phi_0^2\rangle$ at order $\lambda^2$. A dashed line represents the inflaton zero mode and a solid line a matter field. The dot represent the insertion of $\Delta\phi_0^2$. Note that in the $in$-$in$ formalism each vertex can be of two types (not shown.)}\label{fig:variance} 
\end{figure}
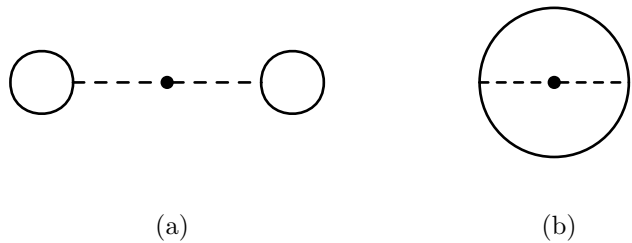

On the other hand, the free operator $\Delta\phi^2$ has   non-zero matrix elements between two states with a single inflaton quantum, or between the vacuum and a state with two inflaton quanta (recall that we do not need to consider vacuum diagrams.) In particular, at quadratic order in $\lambda$, from equation (\ref{eq:vev}),
\begin{subequations}\label{eq:inflaton variance}
\begin{equation}\label{eq:variance}
\langle \Delta\phi^2_V\rangle= \frac{|u|^2}{V}(1+2P_L+2P_3)-\frac{2\sqrt{2} }{V}\mathrm{Im}\left[u^2 T_{LL}\right]
\end{equation}
where $P_L\equiv |T_L|^2$ is the (renormalized) decay probability into a single zero mode,  $P_3$  the decay probability into a pair of matter quanta and a single inflaton, and $T_{LL}$ the transition amplitude into two inflaton quanta.  To arrive at equation (\ref{eq:variance}) we have only included the connected piece of the different matrix elements, as discussed around equation (\ref{eq:disc}).  Looking back at our results for the transition amplitudes and probabilities, one may have expected the variance to grow exponentially with the number of e-folds, but it is easy to check that, in fact,  the leading late time contributions to the variance cancel again. More precisely, using equations (\ref{eq:T1}), (\ref{eq:TL}) and (\ref{eq:TLLb}) we find
\begin{equation}\label{eq:L cancel}
	2\frac{|u|^2}{V}P^\mathrm{ren}_L-\frac{2\sqrt{2}}{V}\mathrm{Im}\, [u^2\,{}^{(c)}T^\mathrm{ren}_{LL}]\approx
	\frac{\lambda^2}{8 \eta^2} \frac{\langle\chi_I^2\rangle_\mathrm{ren}^2}{H^4}\frac{1}{(-k_L t)^{2\eta/3}}.
\end{equation}
Therefore, the exponential growth of $P^\mathrm{ren}_L$ and ${}^{(c)}T^\mathrm{ren}_{LL}$  has no effect  on  the variance of the zero mode.  Incidentally, equation (\ref{eq:L cancel}) is the contribution  to the expectation of $\Delta\phi_V^2$ of  diagram (a) in figure \ref{fig:variance}. As shown in reference \cite{Weinberg:2006ac}, in the $in$-$in$ formalism the expectation value of an observable can be expressed in terms of nested commutators that involve the interaction Hamiltonian and the observable itself. When one expands the nested commutators, some of the resulting expressions are proportional to $[\Delta\phi_V(t_1),\Delta\phi_V(t_2)]$, from which the constant mode in $u$ cancels. This appears to be the origin of the cancellations that we have observed. 
 
There is yet another  cancellation between the two remaining  contributions, which add up to
\begin{equation}\label{eq:lid}
	\frac{2|u|^2}{V}P^\mathrm{ren}_3-\frac{2\sqrt{2}}{V}\mathrm{Im}\, \left[u^2 \,{}^{(d)}T^\mathrm{ren}_{LL}\right]\approx
	\begin{dcases}
	\frac{9\lambda^2}{128\pi^2\eta}\frac{1}{k_L^3 V}\frac{H}{m_\chi}\frac{1}{(-k_L t)^{\eta/3}}
	 & \text{if }H\ll m_\chi,
	\\
	\frac{\lambda^2}{96\pi^2 \eta^2}
	\frac{1}{k_{IR}^3 V}\frac{1}{(-k_L t)^{2\eta/3}} & \text{if } m_\chi\ll H
	\end{dcases}
\end{equation}
\end{subequations}
As  alluded to earlier, we need the subtracted probability $P^\mathrm{ren}_3$ because the expectation of $\Delta\phi_V^2$, which depends on ${}^{(d)}T_{LL}$,  requires renormalization.  By the way, the difference in equation (\ref{eq:lid}) is the contribution of diagram (b) on figure \ref{fig:variance} in the $in$-$in$ formalism.

\subsection{Energy-Momentum Tensor}

Our previous methods  can be also employed to calculate the expectation of the energy-momentum tensor of matter.  Actually, since $\chi$ couples to the inflaton, it is not possible to separate the energy-momentum tensor of $\chi$ from that of the inflaton. Their combined energy-momentum tensor is
\begin{equation}
T_{\mu\nu}=\partial_\mu \phi \partial_\nu \phi
+\partial_\mu \chi \partial_\nu \chi
-\frac{1}{2}g_{\mu\nu}\left(\partial_\rho \phi \partial^\rho \phi+ \partial_\rho \chi \partial^\rho \chi +2V(\phi)+m_0^2 \chi^2+\lambda \phi \chi^2\right),
\end{equation}
where we have included a bare quadratic mass term for the field $\chi$ for later purposes, and $\phi\equiv\bar{\phi}+\Delta\phi$.  Inserting this expansion into the inflaton potential we get
\begin{equation}
V(\phi)=V(\bar{\phi})+V_{\phi}(\bar{\phi})\Delta\phi+\frac{1}{2}V_{\phi\phi}(\bar{\phi})\,\Delta\phi^2+\cdots,
\end{equation}
which is exact for a quadratic potential.  Our goal now is to calculate $\langle T_{\mu\nu}\rangle$ to quadratic order in $\lambda$, by regarding the mass term $\lambda \bar{\phi}\chi^2$ as part of the free theory. 

\paragraph{At zeroth order ($\lambda^0$)} in the interaction there is a contribution to $\langle T_{\mu\nu}\rangle$ stemming from that of the scalar $\phi$ in de Sitter. This one would be present even in the absence of inflaton decays, so we shall ignore it here. There is also a non-perturbative contribution from a free scalar $\chi$ of mass $m_\chi^2=\lambda \bar\phi$ in de Sitter. We  already only calculated the (renormalized) expectation of $\chi_I^2$ using equation (\ref{eq:chiren}), but in this case we can directly borrow the desired result  from the literature (see equation (6.183) in \cite{Birrell:1982ix}),
\begin{equation}
\langle T_{\mu\nu}\rangle^\mathrm{ren}=\frac{g_{\mu\nu}}{64\pi^2}
\left\{m_\chi^2\left(m_\chi^2-2H^2\right)
\left[\psi\left(3/2+\nu\right)
+\psi\left(3/2-\nu\right)
-2\log \frac{m_\chi}{H}\right]
+\frac{4}{3}m_\chi^2 H^2-\frac{29}{15}H^4\right\},
\end{equation} 
where, again, $\psi$ is the digamma function.  Although this applies to a free scalar $\chi$, it depends on $\lambda$ because its mass (\ref{eq:meff}) arises from its interactions with the inflaton. The expectation value is at most of order $H^4$,
which represents a negligible correction to the inflaton energy density as long as $H$ is sub-Planckian.  This is the only correction that does not depend on the normalization of the inflaton zero mode $u$, although it does depend on the value of $\bar{\phi}$.

\paragraph{At first order ($\lambda^1$)} there is a contribution from  the  terms linear in $\Delta\phi$ in the energy-momentum tensor. Because of slow-roll, the non-derivative ones ought to give the dominant contribution,
\begin{equation}\label{eq:delta EMT 1}
	\langle T_{\mu\nu}\rangle^\mathrm{ren} \supset -	V_\phi(\bar{\phi})\,\langle \Delta\phi\rangle^\mathrm{ren}g_{\mu\nu}.
\end{equation}
The renormalized expectation of $\langle \Delta\phi\rangle^\mathrm{ren}$ at first order is quoted in equation (\ref{eq:delta phi}).

\paragraph{At second  order ($\lambda^2$)} the number of terms proliferates significantly. There is a contribution  from the expectation value of the cubic term $\lambda\Delta\phi\chi^2$. This expectation is determined by the transition amplitudes into a single inflaton zero mode $T_L$, and into a zero mode plus two matter quanta, $T_{L\vec{k}-\vec{k}}$.  The first contribution simply reduces to   
\begin{equation}
	\langle T_{\mu\nu}\rangle^\mathrm{ren}_L
	=-\frac{\lambda}{2} g_{\mu\nu} \langle\Delta\phi \rangle^\mathrm{ren}\langle\chi_I^2\rangle^\mathrm{ren},
\end{equation}
where $\langle\Delta\phi \rangle^\mathrm{ren}$ is again that in equation (\ref{eq:delta phi}) and $\langle\chi_I^2\rangle^\mathrm{ren}$ is the expectation of $\chi^2$ at zeroth order, equation (\ref{eq:TL}). Similarly, the contribution of  the second transition to the expectation value of $\lambda\Delta\phi\chi^2$ is 
\begin{equation}\label{eq:T vev 3}
	\langle{T_{\mu\nu}}(t,\vec{x})\rangle_{L\vec{k}-\vec{k}}\approx
	\frac{\lambda g_{\mu\nu}}{V^{3/2}} \sum_{\vec{k}} \,\mathrm{Im} \left[u\, w_k^2\,  T_{L\vec{k}-\vec{k}}\right] .
\end{equation}
Because the integral over momenta in equation (\ref{eq:T vev 3}) logarithmically diverges  in the ultraviolet,  we need to subtract the zeroth order adiabatic approximation to render the integral finite.   In the limit of heavy fields the ensuing integral over $\vec{k}$ can be evaluated exactly and happens to be purely imaginary.  For light fields the correction diverges in the infrared, which is dominated by the contribution of the exact modes, which is the only one we keep.  We thus arrive at 
\begin{equation}
	\langle{T_{\mu\nu}}\rangle^\mathrm{ren}_{L\vec{k}-\vec{k}}\approx
	g_{\mu\nu}\times 
	\begin{dcases}
	\frac{9}{256\pi^2}
	 \frac{1}{k_L^3V} \frac{\lambda^2}{m_\chi^2} H^4
	 \, & \text{if } H\ll m_\chi,\\
	 \frac{1}{48\pi^2 \eta}\frac{1}{k_{IR}^3V}
	 \frac{\lambda^2 H^2 }{(-k_L t)^{\eta/3}}
	 & \text{if } m_\chi \ll H.
	 \end{dcases}
\end{equation}
 
\begin{figure}
\subfigure[]
{
\begin{fmfgraph}(25,25) 
\fmfleft{vl} 
\fmfdot{vl}
\fmfright{vr}
\fmf{plain,left,tension=0.3}{vl,vm,vl}
\fmf{dashes}{vm,vr}
\fmf{plain,tension=0.4}{vr,vr}
\end{fmfgraph}
}
\hspace{3cm}
\subfigure[] 
{
\begin{fmfgraph}(25,25) 
\fmfleft{vl} 
\fmfdot{vl}
\fmfright{vt,vb}
\fmf{plain}{vl,vt}
\fmf{plain}{vl,vb}
\fmf{dashes}{vt,vb}
\fmf{plain,right,tension=0.5}{vt,vb}
\end{fmfgraph}
}
\caption{Contributions to the expectation of $\chi^2$ to order $\lambda^2$ in the $in$-$in$ formalism. The dot represents the insertion of $\chi^2(t)$.}\label{fig:chi squared} 
\end{figure}
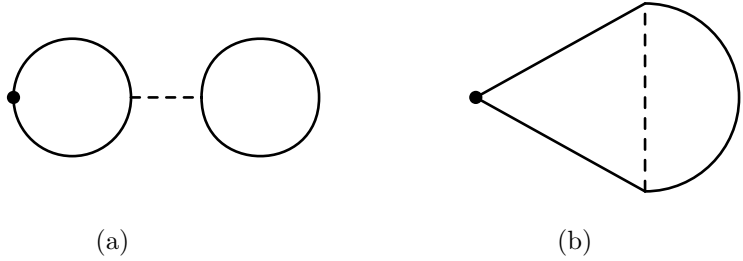

There are additional contributions from the expectation of the quadratic terms in the energy-momentum tensor.  We have already carried out some of the hard work to evaluate these contributions,   because they can be readily calculated from the different transition amplitudes that we have found in section \ref{sec:Transition Amplitudes}. But at this point the calculation becomes increasingly difficult, and we are likely to meet the limits of the adiabatic renormalization scheme. To  gauge the contribution of these quadratic terms we shall limit ourselves to the simplest one, namely, that proportional to the ``bare" mass $m_0^2$. Since the terms with derivatives are accompanied by additional factors of $a^{-2}$, the former is expected to be the fastest growing. Using equation (\ref{eq:vev}) and taking into account the possible decay channels up to second order we get
\begin{multline}\label{eq:T vev 2}
	\langle T_{\mu\nu}\rangle\supset -\frac{m_0^2}{2}\langle\chi^2\rangle g_{\mu\nu}=-\frac{m_0^2}{2}g_{\mu\nu}\Bigg[
	\mathrm{Re} \left(\sum_{\vec{k}} T_L^* <L|\chi_I^2|L \vec{k} -\vec{k}\rangle T_{L \vec{k} -\vec{k}}\right)\\
	-\mathrm{Im}\, \left(\sum_{\vec{k}}\langle 0| \chi_I^2|\vec{k} -\vec{k}\rangle T_{\vec{k}-\vec{k}}\right)
	+\frac{1}{2}\sum_{\vec{k}} \langle L \vec{k} -\vec{k}|\chi_I^2|L \vec{k} -\vec{k}\rangle |T_{L\vec{k}-\vec{k}}|^2\Bigg].
\end{multline} 
Each term in this sum has an interpretation in terms of  diagrams in the $in$-$in$ formalism. Say, the first term  corresponds to the diagram (a) in figure \ref{fig:chi squared},  the second to diagrams (a) and    (b), and the third to diagram (b). The figure does not label the vertices, but recall that in the $in$-$in$ formalism they are of two types. Hence, each diagram gives rise to several mode sums, and thus the multiple correspondence.
 
The mode sums that contain $T_L^* \,T_{L\vec{k}-\vec{k}}$ and ${}^{(b)}T_{\vec{k}-\vec{k}}$ in equation (\ref{eq:T vev 2}) diverge  in the ultraviolet, and it suffices to subtract the adiabatic modes of zeroth adiabatic order to render them finite. Following the adiabatic prescription, we also subtract from the  remaining (finite) mode sums the appropriate zeroth order approximations. In addition, since the first term in equation (\ref{eq:T vev 2})  contains a divergent tadpole  subdiagram, it  appears reasonable to replace the latter by its renormalized counterpart, namely,  the renormalized expectation of $\chi^2$ at zeroth order. In that case, the expectation value in the heavy field limit becomes 
\begin{subequations}\label{eq:chisq 2}
\begin{equation}\label{eq:chisq 2 heavy}
	\langle\chi^2\rangle^\mathrm{ren} \approx
	\frac{87 \lambda^2}{10240\pi^4 \eta} \left(\frac{H}{m_\chi}\right)^4	
	\frac{1}{(-k_L t)^{\eta/3}}
	\quad \text{if } H\ll m_\chi,
\end{equation}
which displays the characteristic suppression by powers of $H/m_\chi$, and a slow grow with the number of e-folds, unlike the transition probabilities it depends on.  If we had not subtracted the zeroth order adiabatic approximation from the finite mode sums, the final results would have been proportional to $(H/m_\chi)^2$ instead.

In the limit of light fields, on the other hand, the three mode sums in equation (\ref{eq:T vev 2}) diverge in the infrared when we approximate them by an integral. Concentrating on the infrared contribution,  neglecting adiabatic subtraction, and keeping only the dominant term in the light field limit we find
\begin{equation}\label{eq:chisq 2 light}
	\langle\chi^2\rangle^\mathrm{ren}\approx	\frac{\lambda^2}{128\pi^4\eta^2}\left(\frac{k_L}{k_{IR}}\right)^3
	\frac{H^2}{m_\chi^2}
	\frac{1}{(-k_L t)^{2\eta/3}}
	\quad  \text{if }  m_\chi \ll H.
\end{equation}
\end{subequations}
Again, there is a secular growth in the expectation value, and the latter is further enhanced by the large factor $(H/m_\chi)^2$. This can only have an impact on the energy-momentum tensor of matter for non-vanishing $m_0$. But, of course, given the symmetries of the theory, there is no reason for  $m_0$ to vanish. As we mentioned, one way to rule out a mass term is to assume that $\chi$ is a Goldstone boson. But in that case, its couplings to the inflaton would need to involve derivatives. 

Finally, it is also instructive to check whether these results have any impact on the energy-momentum tensor of the inflaton field itself. Because we are only concerned with the zero mode, and the latter slowly evolves during inflation, it suffices to consider the expectation of the non-derivative terms, namely,
\begin{equation}
	\langle T_{\mu\nu}\rangle \supset -\frac{V_{\phi\phi}(\bar{\phi})}{2} \langle \Delta\phi_V^2\rangle g_{\mu\nu},
\end{equation} 
in full analogy with equation (\ref{eq:T vev 2}). We already have calculated $\langle \Delta\phi_V^2\rangle$ in equations (\ref{eq:inflaton variance}). Inspection of those equations reveals a moderate impact on the variance of $\phi_V$, in the sense that there is no  exponential growth in either mass limit.

\section{Inflaton Dynamics}

Up to this point we have seen that the inflaton decay probability  rapidly grows during inflation, but that  such a growth does not directly impact the expectation values of the different field operators that we have studied, which in the limit $\eta\to0$ grow with the logarithm of the scale factor.  But if we are interested in knowing whether it is a good approximation to assume that the inflaton  evolves as in the absence of matter couplings, an approach that directly focuses on the evolution of the zero mode  is somewhat more efficient.

\subsection{Quantum Corrected Equation of Motion}

The approach most widely used in the literature to study the impact of quantum corrections on the evolution of the inflation  involves the quantum effective action $\Gamma_\mathrm{eff}$, and the effective evolution equation $\delta \Gamma_\mathrm{eff}/\delta \phi=0$.  Because we are interested in the $in$-$in$ expectation value of the field $\phi$, in order to follow this venue one needs to work with the effective action in the $in$-$in$ formalism, which makes the whole procedure fairly cumbersome. But this procedure is rather heavy-handed anyway. If one is interested in the evolution of  the expectation value of the inflaton it suffices to consider  the Heisenberg equations of motion
\begin{equation}
	i\frac{d\langle \mathcal{O}\rangle}{dt}=
	\langle [\mathcal{O},\mathscr{H}]\rangle,
\end{equation}
where we have assumed that the operator $\mathcal{O}$ does not depend explicitly on time. Focusing on the zero mode of the inflaton and its conjugate momentum, and using the Hamiltonian (\ref{eq:H zero mode}) we thus get
\begin{subequations}\label{eq:Heisenberg}
\begin{align}
 \frac{d\langle\phi_0\rangle}{dt}&=\frac{\langle \pi^\phi_0\rangle}{a^2},\\
	\frac{d\langle\pi_0^\phi\rangle}{dt}&=-m_\phi^2 a^4 \langle \phi_0\rangle-\left\langle \frac{\partial \mathscr{H}_I}{\partial \phi_0}\right\rangle,
\end{align}
\end{subequations}
where we have split the Hamiltonian  into a free piece $\mathscr{H}_0$ (\ref{eq:H0}) and an interaction $\mathscr{H}_I$, ${\mathscr{H}=\mathscr{H}_0+
\mathscr{H}_I}$. Combining both equations in (\ref{eq:Heisenberg}) and using (\ref{eq:zero mode}) we get the ``quantum-corrected" equation of motion
\begin{equation}\label{eq:quantum corrected eom}
	\langle \ddot \phi\rangle+2\mathcal{H} \langle \dot\phi\rangle +m_\phi^2a^2 \langle \phi\rangle+\frac{1}{a^2 \sqrt{V}}\left\langle \frac{\partial \mathscr{H}_I}{\partial \phi_0}\right\rangle=0.
\end{equation}
Note that translational invariance implies that the expectation value of the non-zero modes of $\phi$ vanishes, which is why we can focus on the evolution of the zero mode.  The term in the corrected equation of motion that does not contain time derivatives can be thought of as the derivative of the effective potential in an expanding universe. In particular, note that $V^{-1/2} \partial/\partial\phi_0\equiv \partial/\partial\phi_V$. If the inflaton potential is not quadratic, the quantum corrected equation of motion still has the form (\ref{eq:quantum corrected eom}), with $\langle\phi\rangle$ replaced by $\langle \Delta\phi\rangle$,  $m^2_\phi$ by $V_{\phi\phi}(\bar{\phi})$ and $\partial/\partial \phi_0$ by  $\partial/\partial \Delta\phi_0$.

Clearly, the expectation value of $\phi$ obeys the classical equation of motion, modulo  corrections given by the  expectation of $\partial \mathscr{H}_I/\partial \phi_0$. For the interaction Hamiltonian (\ref{eq:H I}), in particular,
\begin{equation}\label{eq:quantum correction}
\frac{1}{a^2\sqrt{V}}\left\langle\frac{\partial\mathscr{H}_I}{\partial\phi_0}\right\rangle=\frac{a^2 \lambda}{2}
\langle \chi^2(t,\vec{x})\rangle.
\end{equation}
By construction, these quantum corrections  are real, since the operator $\partial \mathscr{H}_I/\partial \phi_0$ is hermitian. In the light of the last equation, the quantum corrected equation of motion has a natural and simple interpretation: For arbitrary values of $\chi$, the classical equation of motion of the homogeneous scalar $\phi$ has the form
$	\ddot{\phi}+2\mathcal{H}\dot{\phi}+m_\phi^2 a^2 \phi+\frac{\lambda}{2}a^2\chi^2=0.
$ 
When $\chi$ is in the vacuum state, we assign to $\chi$ the classical value $\chi=0$, and the previous equation  reduces to that of the classical background field $\bar{\phi}$. But quantum-mechanically, we cannot set $\chi$ to zero, since it experiences vacuum fluctuations. The correction term in (\ref{eq:quantum corrected eom}) is simply what we get when we replace $\chi^2$ by its vacuum expectation value $\langle \chi^2\rangle$. Note that the quantum-corrected equation of motion does not have the form that has been often quoted in the literature \cite{Kofman:1997yn,Allahverdi:2010xz},
$
\langle \ddot{\phi}\rangle +(2\mathcal{H}+\Gamma a)\langle \dot{\phi}\rangle+m_\phi^2 a^2 \langle \phi\rangle=0.
$
The latter contains an additional damping term  proportional to $\langle \dot{\phi}\rangle$ that does not appear in (\ref{eq:quantum corrected eom}). In particular, as we shall see, in de Sitter space quantum corrections simply add an additional  constant driving force to the equation of motion of the inflaton at first order. 
 
The evaluation of $\langle  \partial \mathscr{H}_I/\partial\phi_0\rangle$ basically amounts to the calculation of $\langle \chi^2(t,\vec{x})\rangle$, which is one of the main focuses of quantum field theory in curved spacetimes. In fact, there  also is a nice parallel between the equation of motion (\ref{eq:quantum corrected eom}) and the equations of semiclassical gravity in which such calculations are carried out.  In the latter the gravitational field is sourced by the expectation value of the energy-momentum tensor, $G_{\mu\nu}=8\pi G \,\langle T_{\mu\nu}\rangle$. These semiclassical equations are the gravitational (non-linear) analogues  of equation (\ref{eq:quantum corrected eom}), with the classical metric playing the role of $\langle\phi\rangle$, and the energy momentum tensor playing the role of $\partial \mathscr{H}_I/\partial \phi_0$. 

In the cases we have analyzed, the expectation value of $\chi^2$ grows with a power of time, ${\langle\chi^2\rangle\propto(-t)^p}$. From equations (\ref{eq:int diff heavy}) and (\ref{eq:int diff light}), $p=0$ at order $\lambda^0$, and from equations (\ref{eq:chisq 2})  $p=-\eta/3$ or $p=-2\eta/3$ at order $\lambda^2$.  A particular solution of (\ref{eq:quantum corrected eom}) in those instances is
\begin{equation}\label{eq:delta phi dynamical}
\Delta\bar\phi=-\frac{\lambda}{2}
	\frac{\langle\chi^2\rangle}{m_\phi^2+p(p-3)H^2},
\end{equation}
which can be thought of as the correction to the inflaton background value due to quantum effects.  Quantum corrections are  negligible whenever
$\Delta\bar{\phi}\ll\bar{\phi}$. This condition in some sense replaces the condition that the term proportional to $\lambda$ in the effective potential (\ref{eq:Veff}) be subdominant. But comparison of both expressions shows that they are in fact very different in nature. Such a disagreement suggests that in some cases it may not be justified to apply quantum corrections derived in Minkowski spacetime to field theories in an expanding universe.

If we had tried to calculate  $\langle \Delta\phi_V\rangle$ to third order in $\lambda$ within the $in$-$in$ formalism, we would have had to evaluate a relatively complicated expression containing three time integrals of the expectation of a term cubic in the interaction. In the present approach, since $\partial\mathscr{H}_I/\partial \phi_0$ is already proportional to $\lambda$, it suffices to calculate $\langle \chi^2\rangle$ to second order, which considerably simplifies the analysis. 

\section{Summary and Conclusions}

In most inflationary scenarios the universe inflates until the inflaton reaches the vicinity of the bottom of its potential, where the violation of the slow-roll conditions triggers the end of inflation. It is  after such an end that the inflaton is supposed to decay into matter  and thus reheat the universe. 

But in order for the previous picture to hold, the inflaton must survive  the inflationary stage.    The condition that is taken to signal the decay of inflaton after the end of inflation is $\Gamma H^{-1}\gg 1$, where $\Gamma$ is the decay rate of the inflaton and $H$ the Hubble constant. Since $H^{-1}$ is proportional to cosmic time, this is equivalent to the demand that the total  decay probability of the inflaton become large. Therefore, we would expect the inflaton to survive until the end of inflation as long as its total decay probability remains small. 

In Section \ref{sec:Transition Amplitudes} we have calculated various decay probabilities of the inflaton during inflation. We have seen that these  probabilities  grow rapidly, apparently implying that the inflaton should decay just after a relatively small number of e-folds.  But closer inspection reveals that this growth is not  translated into exponentially large corrections to the expectation value of the inflaton or the energy-momentum tensor of its decay products, because there are cancellations among the different terms that contribute to the expectation values.  Since we are dealing with a field theory anyway (and are not interested in $S$-matrix elements) evaluation of the expectation of different field operators appears  to be a better strategy to discern whether the inflaton is effectively decaying during inflation.

As an illustration, we shall begin by looking at the  impact of such decays on the expectation  of the inflaton itself.  From equation (\ref{eq:delta phi}), and because $m_\chi^2=m_0^2+\lambda\bar{\phi}\geq \lambda\bar\phi$, the leading correction is bounded by the model-independent limit
\begin{equation}\label{eq:limit1}
	\left|\frac{\Delta\bar\phi}{\bar\phi}\right|\lesssim\frac{3}{16\pi^2} \frac{1}{\eta}\frac{H^2}{\bar{\phi}^2}\frac{1}{(-k_L t)^{2\eta/3}}\ll 1,
\end{equation}
which interestingly,  does not depend on the coupling constant $\lambda$.  Recall that $\eta\equiv V_{\phi\phi}/H^2$ is a slow-roll parameter, and that $1/(-k_L t)$ equals $e^{N_L}$, where $N_L$ is the total number of e-folds of inflation. Because in any reasonable inflationary model the size of the field's quantum fluctuations ought to be much smaller than the field itself ($H\ll \bar{\phi}$), at least while observable scales are exiting the horizon,  we expect the impact of the decay on the background field to be small, unless the number of e-folds is very large, $N_L\eta\gg 1$.  On the other hand, these corrections could play an important role during self-reproduction in hilltop inflationary models \cite{Barenboim:2016mmw}, in which eternal inflation occurs at field values $H/\bar{\phi}\gtrsim \eta$.  We obtain a similar constraint using the dynamical correction (\ref{eq:delta phi dynamical}) at first order in $\lambda$. The demand that the order $\lambda$ correction to the energy density (\ref{eq:delta EMT 1}) be smaller than that of the background also leads to a similar, albeit weaker, limit. 

In order to bound the value of the coupling constant $\lambda$ we need to consider corrections at order $\lambda^2$. To avoid an excessive proliferation of parameters, let us assume that $m_0^2$ is negligible. Then, the only relevant contributions at second order are those to $\langle \Delta\phi_V^2\rangle.$ The quantum corrections that we have calculated are proportional to different positive powers of $H/m_\chi$, and are thus expected to be tighter for light matter fields. Demanding that the light field limit correction to the energy density associated with (\ref{eq:lid})  be smaller than  that of the background we obtain
\begin{subequations}\label{eq:lambda limit}
\begin{equation}\label{eq:lambda limit light}
	\left(\frac{\lambda}{\bar{\phi}}\right)^2\ll \mathrm{min}\left\{(H/\bar{\phi})^4,96\pi^2 \eta^2 \,(-k_L t)^{2\eta/3}\right\}.
\end{equation}
There is an additional condition here because $\lambda$ needs to be small enough for the field to be light, $\lambda\bar{\phi}=m_\chi^2\ll H^2$.
Equation (\ref{eq:lambda limit light}) is thus not very constraining because it can be satisfied for small enough $\lambda$.   Note that $H/\bar{\phi}$ is typically very small, particularly in chaotic inflationary models. Therefore, for light matter fields quantum corrections are  expected to be automatically small.    In the heavy field limit the same analysis of equation (\ref{eq:lid}) returns the conditions
 \begin{equation}\label{eq:lambda limit heavy}
 \left(\frac{H}{\bar{\phi}}\right)^4\ll \left(\frac{\lambda}{\bar{\phi}}\right)^2\ll
 \left(\frac{128 \pi^2 }{9} 
 \frac{\bar\phi}{H}(-k_L t)^{\eta/3} \eta\right)^{4/3},
 \end{equation}
\end{subequations}
meaning that the coupling constant $\lambda$ has to be large enough for the matter field to be heavy, but not large enough for it to decay too rapidly.  The upper limit in equation (\ref{eq:lambda limit heavy}) is again not very restrictive because we expect $\bar{\phi}/H$ to be  much larger than the other parameters.  We could derive various similar constraints by applying the same methods to other  expectation values, but these would not produce anything significantly different. We just note that since many of the quantum corrections are proportional to different powers of $e^{\eta N_L}$, there appears to be an upper limit on the number of e-folds of inflation of the order $N_L\sim 1/\eta$.  Alternatively, in the limit $N_L\eta\leq 1$ our results are not very sensitive to the normalization of the inflaton zero mode fluctuations, which is determined by $k_L$. In the limit $\eta\to 0$, the different corrections grow with the logarithm of the scale factor, as many other loop corrections to inflationary observables.

Incidentally, note that if we use the criterion $\Gamma H^{-1}\ll 1$ as a proxy for the survival of the inflaton during inflation, where $\Gamma=\lambda^2/(32\pi m_\phi)$ is the decay rate of the inflaton in flat spacetime,   we obtain a criterion that
 is very different from those in equations (\ref{eq:lambda limit}), especially because the former does not involve the background field $\bar{\phi}$. On the other hand,  if $\Gamma H^{-1}\gtrsim 1$ does signal the decay of the inflaton \emph{after} the end of inflation,  equations (\ref{eq:lambda limit})  imply that there is a wide range of coupling constants for which the inflaton decays shortly after the end of inflation,  but not during it. More precisely, the inflaton will typically decay long after inflation if the matter fields it couples to remain light during inflation, but it will decay shortly after the end of inflation if, on top of equation (\ref{eq:lambda limit heavy}) the coupling constant $\lambda$ satisfies
 \begin{equation}
 32\pi \frac{m_\phi}{\bar{\phi}}\frac{H}{\bar{\phi}}\lesssim\left(\frac{\lambda}{\bar{\phi}}\right)^2.
 \end{equation}

Our results can be also interpreted in a different light. We have argued above that  the zero mode $\phi_{\vec{k}=0}$ can be thought of as a proxy for the longest mode that left the horizon during inflation. Therefore,  they suggest that the inflaton couplings responsible for its decay cannot alter  the power spectrum of the inflaton on the largest scales significantly, as long as $N_L \eta\leq 1$.  This result agrees with the    the conclusions of references \cite{Weinberg:2005vy,Weinberg:2006ac}, even though  the interaction responsible for the inflaton decay is not one of the ``safe" or ``dangerous" interactions discussed therein. 

Finally let us stress that since our conclusions mostly involve renormalized quantities, they  depend on the validity of the adiabatic scheme for regularization and renormalization.  This is why it  would be to useful to repeat our analysis with a more rigorous  renormalization scheme, although at this point there does not appear to be a clear consensus as to what the latter should be. Nevertheless, although  adiabatic subtraction suppresses quantum corrections by additional factors of $H/m_\chi$ in the limit of heavy fields, it does not have much of an impact in the opposite limit. 
 
It is fair to say that reheating after  inflation remains one of the least investigated aspects of inflation, in spite of the already significant amount of literature devoted to the topic. Yet the impact on inflation from the couplings necessary for reheating  essentially remained unexplored. In this work we have barely scratched the surface of the subject by analyzing one of the simplest decay-inducing interactions.  In this simple case we have encountered potentially large corrections  in the limit $N_L\eta\gg 1$, which could signal large corrections to the  primordial spectrum of scalar perturbations on large scales. Because of this possibility alone, we believe that the topic deserves further scrutiny.

\begin{appendix}

\section{Adiabatic Modes}
\label{sec:Adiabatic Modes}

In the adiabatic subtraction scheme one needs to subtract from a divergent expectation value an  adiabatic approximations obtained by replacing the exact mode functions by  adiabatic approximations. The latter are solutions of the mode equation (\ref{eq:w motion}) expanded in powers of an appropriate
slowness parameter. In order to present the form of these solutions, let us first introduce the scaled mode function $\tilde{w}_k\equiv a w_k$, which obeys the  equation
\begin{equation}\label{eq:mode}
	\ddot{\tilde{w}}_k+\left(k^2+m_\chi^2 a^2 -\frac{\ddot{a}}{a}\right) \tilde{w}_k=0.
\end{equation}
This is useful because the equation of  motion for $\tilde{w}_k$ resembles that of an harmonic oscillator with a time-dependent frequency.  The adiabatic approximation is essentially a limit of slow expansion.  To study this limit we replace the scale factor $a(t)$ by $a(\alpha t)$, and consider the limit of small $\alpha$. Inserting this scale factor in the equation (\ref{eq:mode}) and changing variables $\alpha t\to t$ leads to
\begin{equation}
	\ddot{\tilde{w}}_k+\omega^2 \tilde{w}_k=0,\quad
	\text{where}\quad
	\omega^2\equiv \frac{k^2+m_\chi^2 a^2}{\alpha^2} -\frac{\ddot{a}}{a}.
\end{equation}
The (positive frequency) normalized solution of this equation can be written down in WKB form
\begin{equation}\label{eq:WKB}
	\tilde{w}_k(t)=\frac{1}{\sqrt{2W}}\exp\left(-i \int^t W(t_1)  \, dt_1\right),
\end{equation}
where $W$ obeys the relation
\begin{equation}
	\omega^2=W^2+\frac{1}{2}\frac{\ddot{W}}{W}-\frac{3}{4}\frac{\dot W^2}{W^2}.
\end{equation}
This equation can be solved recursively by expanding in powers of $\alpha$. To leading order,
\begin{equation}
	W^{(0)}\equiv \omega_0=\frac{1}{\alpha}\sqrt{k^2+m_\chi^2 a^2},
\end{equation}
which we shall label as zeroth adiabatic order. The term of order $\alpha^0$ vanishes, and that of order $\alpha$ is 
\begin{equation}
	W^{(2)}=\omega_0+\frac{\alpha}{2\omega_0} \left(\frac{3}{4}\frac{\dot{\omega}^2_0}{\omega_0^2}-\frac{1}{2}\frac{\ddot{\omega}_0}{\omega_0}-\frac{\ddot a}{a}\right),
\end{equation}
which we shall label as the second adiabatic order.  Luckily, we shall not need higher orders here. We obtain the adiabatic mode to order $n$ simply by inserting the order $n$ approximation to $W$ into the solution (\ref{eq:WKB}). Therefore, 
\begin{equation}
	w_k^{(n)}=\frac{1}{a}\frac{1}{\sqrt{2 W^{(n)}}}\exp\left(-i \int^t W^{(n)}  \, d\tilde t\right).
\end{equation}
Once we have obtained the expansion of a given quantity to the desired order, we set of course $\alpha=1$. 

\section{Uniform Expansion for Heavy Fields}
\label{sec:Uniform Expansion}

In order to evaluate some of the transition amplitudes and probabilities it is useful to have appropriate approximations for the matter field mode functions in terms of elementary functions. In the limit of heavy fields, these are obtained from  the uniform expansion of the Hankel function \cite{Dunster} 
\begin{equation}\label{eq:Dunster}
	 H_1(i\mu,z)= 
	\sqrt{\frac{2}{\pi}}
	e^{-i\pi/4} e^{\pi\mu/2}
	\frac{e^{i\,\mu\, \xi} }{(\mu^2+z^2)^{1/4}}
	\sum_{s=0}^{n-1} \frac{U_s(p)}{(i\mu)^s}+\mathcal{O}(\mu^{-n}),
\end{equation}
where we have abbreviated
\begin{equation}
	\xi\equiv\sqrt{1+z^2/\mu^2}+\log\left(\frac{z}{\mu+\sqrt{\mu^2+z^2}}\right),  \quad
	p=\frac{1}{\sqrt{1+z^2/\mu^2}},
\end{equation}
and where the functions $U_s(p)$ are those of equation (7.10) in Chapter 10 of reference \cite{Olver}. We shall need to keep terms up to order $\mu^{-5}$ in the expansion at most, so we just gather here the first four functions
\begin{subequations}
\begin{align}
	U_0&=1,\\
	U_1&=\frac{3p-5p^3}{24}, \\
	U_2&=\frac{81 p^2-462 p^4+385p^6}{1152}\\
	U_3&=\frac{30375p^3-369603p^5+765765p^7-425425p^9}{414720},\\
	U_4&=\frac{4465125 p^4-94121676p^6+349922430 p^8-446185740 p^{10}+185910725p^{12}}{39813120}.
\end{align}
\end{subequations}

\end{appendix}
\end{fmffile}

\end{document}